\documentclass[aps, prd, 10pt, notitlepage, superscriptaddress, nofootinbib,numbers,showpacs]{revtex4-2}

\usepackage[utf8]{inputenc}
\usepackage[T1]{fontenc}
\usepackage{anyfontsize}
\usepackage{amsmath}
\usepackage{amssymb}
\usepackage{amsfonts}
\usepackage{mathrsfs}
\usepackage{enumitem}
\usepackage{microtype}
\usepackage[normalem]{ulem}
\usepackage{graphicx}
\usepackage{epsfig}
\usepackage{float}
\usepackage[dvipsnames]{xcolor}
\usepackage{hyperref}
\hypersetup{
    colorlinks=true,
    citecolor=Purple,
    linkcolor=Purple,
    urlcolor=Purple,
    linktocpage=true,
    breaklinks=true
}
\usepackage{soul}
\usepackage[capitalize]{cleveref}

\begin{document}

\title{Traversable Casimir Wormholes with Gravitational Memory}

\author{Jonathan A. Rebouças}
\email{jalvesreboucas@ifce.edu.br}
\affiliation{Instituto Federal de Educação, Ciência e Tecnologia do Ceará (IFCE), Iguatu, Brazil}

\author{Francisco Bento Lustosa}
\email{chico.lustosa@uece.br}
\affiliation{Universidade Estadual do Cear\'a (UECE), Faculdade de Educa\c{c}\~ao, Ci\^encias e Letras de Iguatu, Av. D\'ario Rabelo s/n, Iguatu - CE, 63.500-000 - Brazil.}

\author{Celio R. Muniz}
\email{celio.muniz@uece.br}
\affiliation{Universidade Estadual do Cear\'a (UECE), Faculdade de Educa\c{c}\~ao, Ci\^encias e Letras de Iguatu, Av. D\'ario Rabelo s/n, Iguatu - CE, 63.500-000 - Brazil.}

\date{\today}

\begin{abstract} We investigate a class of traversable wormhole geometries supported by an effective Casimir source corrected by gravitational memory. The construction is motivated by the fact that a time-dependent gravitational background can leave a permanent positive shift in the vacuum polarization of a quantum field confined to a Casimir cavity. By promoting the plate separation to an effective radial scale in the Morris--Thorne spacetime, we obtain a density profile composed of the usual negative Casimir contribution, proportional to $r^{-4}$, and a positive memory-induced correction, proportional to $r^{-7}$. The corresponding shape function is derived directly from the Einstein equations and satisfies the throat condition by construction. We determine the redshift function from a constant barotropic equation of state together with the requirement of regularity at the throat, which fixes the barotropic parameter in terms of the Casimir and memory coefficients. The flare-out condition defines the admissible range of the memory parameter and separates a Casimir-dominated sector from a phantom-like sector, with the transition point associated with a singular limit of the constant-barotropic description. We analyze the curvature scalar, the embedding structure, the energy conditions, and the Tolman--Oppenheimer--Volkoff equilibrium of the anisotropic matter source. The radial null energy condition is necessarily violated at the throat, while the tangential sector depends sensitively on the redshift gradient. We also examine the shadow radius as a phenomenological diagnostic and show that admissible solutions can overlap the Event Horizon Telescope range for M87*. The results indicate that gravitational memory can deform Casimir-supported wormholes by softening the ordinary Casimir contribution, modifying the near-throat geometry, and reshaping the internal stress balance required to sustain traversability. \end{abstract}

\maketitle

\section{Introduction}
\label{sec:introduction}

Wormholes are among the most intriguing solutions of general relativity, since they represent nontrivial spacetime topologies connecting distant regions of the same universe, or even distinct universes, through a finite throat. The historical precursor of this idea is the Einstein--Rosen bridge \cite{EinsteinRosen1935}, although such a construction is not traversable in the modern sense. The systematic formulation of traversable wormholes was established by Morris and Thorne \cite{MorrisThorne1988}, who introduced a static and spherically symmetric geometry characterized by a shape function and a redshift function, together with the geometric conditions required for safe traversal. Shortly afterwards, Morris, Thorne and Yurtsever \cite{MorrisThorneYurtsever1988} showed that traversable wormholes may also lead to chronology-violation mechanisms, emphasizing their deep connection with the causal structure of spacetime.

Since then, wormhole physics has evolved into a broad field. Visser developed spacetime-surgery methods and thin-shell constructions, providing simple examples in which the exotic matter is concentrated at a junction surface \cite{Visser1989}. The stability of such thin-shell configurations was later analyzed by Poisson and Visser \cite{PoissonVisser1995}. Hochberg and Visser clarified the relation between the flare-out condition, trapped or anti-trapped surfaces, and energy-condition violation in both static and dynamical geometries \cite{HochbergVisser1998}. Rotating traversable wormholes were introduced by Teo \cite{Teo1998}, opening the way to more realistic compact-object phenomenology. A persistent conclusion of these developments is that, within standard Einstein gravity, traversable wormholes generically require matter sources violating the null energy condition. Nevertheless, several works have shown that the total amount of exotic matter can be quantified and, in some cases, made arbitrarily small \cite{VisserKarDadhich2003,NandiZhangKumar2004,FewsterRoman2005}. Thus, one of the central questions in wormhole physics is not merely whether exotic matter is required, but whether physically motivated quantum sources can provide it in a controlled way.

A natural candidate for such a source is the Casimir effect. Originally predicted by Casimir for two parallel conducting plates \cite{Casimir1948}, and closely related to the retarded interaction studied by Casimir and Polder \cite{CasimirPolder1948}, the Casimir effect arises from the modification of quantum vacuum modes by boundary conditions. Its stress-energy tensor between conducting plates was derived by Brown and Maclay \cite{BrownMaclay1969}, and the corresponding force has been measured with increasing precision in micrometer and submicrometer experiments \cite{Lamoreaux1997,MohideenRoy1998}. Theoretical developments have since clarified the role of geometry, material properties, finite temperature, regularization, and vacuum stresses, as discussed in the standard reviews by Bordag, Mohideen and Mostepanenko \cite{BordagMohideenMostepanenko2001} and Milton \cite{Milton2004}. From the viewpoint of gravity, the relevant feature is that the Casimir effect provides a negative energy density of quantum origin, making it one of the most physically motivated sources for sustaining wormhole throats.

The gravitational role of Casimir energy has also been extensively investigated. Jaffe \cite{Jaffe2005} discussed the conceptual status of the Casimir effect and its relation to vacuum fluctuations, while Fulling, Milton, Parashar and collaborators analyzed how Casimir energy responds to gravitational fields and inertial effects \cite{FullingMiltonParasharRomeoShajeshWagner2007,MiltonParasharShajeshWagner2007,ShajeshMiltonParasharWagner2008,MiltonShajeshFullingParashar2014}. Their results support the view that Casimir energy gravitates consistently with the equivalence principle when the complete system is properly regularized. Other works studied weak gravitational corrections to the Casimir effect \cite{NazariNouriZonoz2010,Sorge2019}, as well as Casimir energies in freely falling frames and black-hole backgrounds \cite{SorgeWilson2019,WilsonSorgeFulling2020}. These studies are important for the present problem because they reinforce the idea that Casimir stresses can be treated as genuine contributions to the semiclassical stress-energy tensor.

The relation between Casimir physics and wormholes has been explored from two complementary perspectives. In one direction, quantum vacuum effects are computed in a prescribed wormhole background. Khabibullin, Khusnutdinov and Sushkov \cite{KhabibullinKhusnutdinovSushkov2006} studied the Casimir effect in a wormhole spacetime and showed that the vacuum energy may display attractive or repulsive behavior depending on the parameters of the field. Bezerra et al. \cite{BezerraBezerraMelloKhusnutdinovSushkov2010} investigated the vacuum polarization of a massive scalar field in a wormhole geometry and concluded that the induced stress tensor does not generally self-support the throat. In the opposite direction, Casimir-like stresses can be used as explicit sources of the wormhole geometry. Garattini and Lobo \cite{GarattiniLobo2009} analyzed self-sustained wormholes supported by one-loop graviton contributions, while Garattini \cite{Garattini2019} constructed a class of Casimir wormholes in which the Casimir stress is directly responsible for sustaining the geometry. More recently, this program has been extended to Yang--Mills Casimir wormholes in both lower-dimensional and four-dimensional settings \cite{SantosMunizMaluf2023,Santos2024}, to Casimir-supported configurations in modified symmetric teleparallel gravity, including generalized-uncertainty-principle corrections \cite{Hassan2022,Hassan2023}, and to rotating Casimir wormholes \cite{GarattiniTzikas2025}. These works suggest that Casimir-supported wormholes form a promising semiclassical route for connecting exotic spacetime geometries with physically motivated quantum sources.

A second ingredient of the present work comes from gravitational memory. The memory effect refers to the permanent imprint left in spacetime after the passage of gravitational radiation. Its linear form was anticipated by Zel'dovich and Polnarev \cite{ZeldovichPolnarev1974} and by Braginsky and Grishchuk \cite{BraginskyGrishchuk1985}. The nonlinear memory effect was discovered by Christodoulou \cite{Christodoulou1991} and physically interpreted by Thorne \cite{Thorne1992} as a gravitational-wave burst with a permanent displacement effect. In modern language, gravitational memory is deeply connected with the infrared structure of gravity. Weinberg's soft graviton theorem \cite{Weinberg1965}, the BMS asymptotic symmetry group, and gravitational-wave memory were shown to form different aspects of the same infrared structure \cite{HeLysovMitraStrominger2015,StromingerZhiboedov2016,PasterskiStromingerZhiboedov2016}. From the phenomenological side, Favata studied the nonlinear memory produced by binary black-hole mergers and reviewed its detectability \cite{Favata2009,Favata2010}. Further developments include electromagnetic analogues and Einstein--Maxwell contributions to memory \cite{BieriGarfinkle2013,BieriChenYau2012}, the interpretation of memory in terms of null stress-energy and radiation bursts \cite{TolishWald2014}, prospects for detection with gravitational-wave observatories \cite{LaskyThraneLevinBlackmanChen2016}, and extensions beyond general relativity \cite{HeisenbergYunesZosso2023}.

The point of contact between the Casimir program and the gravitational-memory program was recently made explicit by Sorge \cite{Sorge:2023sdf}. In that work, a Casimir cavity is placed in a time-dependent gravitational background with asymptotically Minkowskian regions in the far past and future. Using Schwinger's proper-time method, Sorge showed that the passage of the gravitational perturbation leaves a permanent residual shift in the vacuum polarization of the confined field. In the weak-gravitational-wave limit, this shift appears as a positive correction to the usual negative Casimir energy density, scaling with a higher inverse power of the plate separation. This result is the central physical input of the present work. It suggests that the Casimir vacuum cannot always be regarded as a purely static source: its effective energy density may carry information about the previous gravitational history of the system. Motivated by this observation, we investigate whether such a memory-corrected Casimir density can act as the matter source of a traversable wormhole.

In the present paper, we construct a class of traversable wormhole geometries supported by an effective Casimir source corrected by gravitational memory. The ordinary Casimir contribution provides the negative component required to sustain the throat, while the memory term introduces a positive correction associated with the past gravitational perturbation. By promoting the plate separation scale of the Casimir setup to an effective radial scale in a Morris--Thorne geometry, we obtain a density profile containing a standard $r^{-4}$ Casimir term and a memory-induced $r^{-7}$ correction. We then derive the corresponding shape function, determine the redshift sector through a barotropic relation and throat regularity, and analyze the parameter domain compatible with the flare-out condition and asymptotic flatness. The main objective is to understand how the gravitational memory of the quantum vacuum modifies the geometry, the matter sector, and the physical viability of a Casimir-supported wormhole.

This paper is organized as follows. In Sec.~\ref{sec:model}, we introduce the Casimir source with gravitational memory and review the geometric framework of traversable Lorentzian wormholes. In Sec.~\ref{sec:geometry}, we derive the shape function, analyze the redshift function, discuss the geometric restrictions on the parameters, and study curvature and embedding properties. In Sec.~\ref{sec:matter}, we examine the matter distribution, the energy conditions, and the equilibrium condition through the Tolman--Oppenheimer--Volkoff equation. In Sec.~\ref{sec:phenomenology}, we discuss the shadow of the wormhole as a possible phenomenological diagnostic. Finally, Sec.~\ref{sec:conclusion} summarizes our main results and presents the conclusions.

\section{Casimir-supported traversable wormholes with gravitational memory}
\label{sec:model}

In this section, we introduce the physical ingredients of our construction. We first summarize the gravitational-memory correction to the Casimir effect and then translate its leading behavior into an effective radial density profile. After that, we review the Morris--Thorne geometry and the basic constraints required for a Lorentzian traversable wormhole.

\subsection{Casimir effect with gravitational memory}
\label{subsec:sorge}

The standard Casimir effect arises from the modification of quantum vacuum modes by boundary conditions. For two parallel plates separated by a proper distance $L$, the electromagnetic Casimir energy density in flat spacetime is
\begin{equation}
\rho_{\rm Cas}(L)=-\frac{\pi^2}{720L^4}.
\label{eq:casimir_standard}
\end{equation}
The negative sign is essential for the present purpose, since traversable wormholes require a violation of the NEC at or near the throat.

The analysis of Sorge \cite{Sorge:2023sdf} shows that a time-dependent gravitational background may leave a permanent imprint on the vacuum polarization inside a Casimir cavity. In the weak-field regime, after the gravitational perturbation has passed, the vacuum energy receives a positive residual correction. For a Gaussian gravitational pulse, the leading correction has the form
\begin{equation}
\delta\rho_{\rm mem}(L)\simeq
\frac{15H^2}{64\sqrt{2\pi}\,\sigma^3L^7},
\label{eq:sorge_scalar}
\end{equation}
in natural units, for a scalar field. Here $H$ is the strain amplitude of the gravitational perturbation and $\sigma^{-1}$ characterizes its duration. For the electromagnetic case, the result is doubled because of the two polarization states. Therefore, the memory contribution is not another negative-energy term; rather, it is a positive correction that weakens the magnitude of the ordinary Casimir energy.

This feature is central to our model. The wormhole is not supported by a purely standard Casimir density, but by a Casimir-like density whose strength has been modified by the previous gravitational history of the vacuum. The memory correction will therefore affect not only the intensity of the exotic matter but also the geometric conditions that determine whether a traversable throat can exist.

\subsection{Effective density profile}
\label{subsec:density}

We now promote the plate separation scale $L$ to an effective radial scale $r$. This gives the phenomenological source
\begin{equation}
\rho(r)=-\frac{\alpha}{r^4}+\frac{\eta}{r^7}
       =\frac{\eta-\alpha r^3}{r^7},
\label{eq:density_profile}
\end{equation}
where $\alpha$ controls the ordinary Casimir contribution and $\eta$ controls the gravitational-memory correction. The first basic restrictions are $\alpha>0$ and $\eta\geq 0$. The condition $\alpha>0$ ensures that the first term in Eq.~\eqref{eq:density_profile} is negative, while $\eta\geq0$ follows from the positive nature of the memory correction.

For an electromagnetic Casimir source one may set $\alpha=\pi^2/720$, while for a scalar-field model $\alpha=\pi^2/1440$. The same distinction applies to the memory coefficient. In the scalar case, Eq.~\eqref{eq:sorge_scalar} gives $\eta=15H^2/(64\sqrt{2\pi}\sigma^3)$, while the electromagnetic case gives twice this value. In the original Casimir-memory calculation, this contribution is obtained as a perturbative correction to the ordinary Casimir energy. In the present wormhole construction, however, $\eta$ will be treated as an effective parameter controlling the relative strength of the memory-induced contribution in the radial source. This allows us to distinguish the conservative sector, $\eta/(\alpha r_0^3)\ll1$, where the memory term remains a small correction to the ordinary Casimir density at the throat, from an extended effective sector in which the memory correction becomes comparable to the Casimir scale. The latter should be understood as a phenomenological continuation of the leading-order density profile, useful for probing how the geometry responds when the memory contribution becomes structurally relevant.

The density changes sign at $r_{\rm c}^{\,3}=\eta/\alpha$. Thus, for $r^3>\eta/\alpha$ the ordinary Casimir contribution dominates and $\rho(r)<0$, while for $r^3<\eta/\alpha$ the memory-induced term dominates and $\rho(r)>0$. In a wormhole geometry the most restrictive comparison is made at the throat scale $r_0$. If $\eta<\alpha r_0^3$, the density at the throat remains negative and the source preserves the standard Casimir character in the whole exterior region. If $\eta>\alpha r_0^3$, the near-throat density becomes positive, although the ordinary Casimir term eventually dominates again at sufficiently large $r$ because the memory contribution decays faster. This transition is not, by itself, a geometric viability condition. Rather, it identifies the point at which the effective source leaves the strictly Casimir-dominated sector and enters a phenomenological regime where the memory correction has a leading influence on the near-throat matter content.

\begin{figure}[!htp]
\centering
\includegraphics[width=0.6\textwidth]{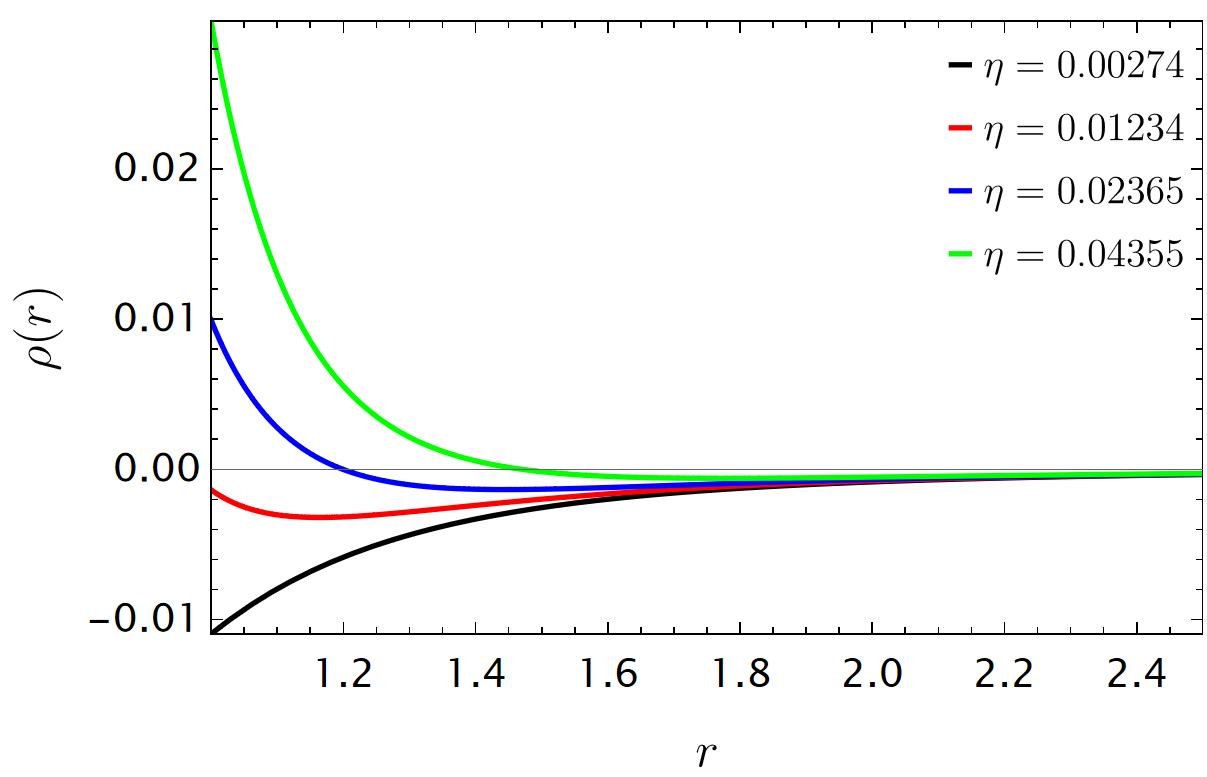}
\caption{Effective Casimir density with gravitational memory, $\rho(r)$, for different values of the memory parameter $\eta$ as a function of the radial coordinate $r$, with $\alpha=\pi^2/720$ and $r_0=1$ fixed. The curves correspond to $\eta=0.00274$, $0.01234$, $0.02365$, and $0.04355$. The horizontal line marks $\rho(r)=0$, showing the transition between negative Casimir-dominated regions and positive memory-dominated regions.}
\label{fig:density_profile}
\end{figure}

Figure~\ref{fig:density_profile} shows the behavior of the effective density profile generated by the competition between the standard Casimir contribution and the gravitational-memory correction. For the smaller values of $\eta$, the density remains negative throughout the interval shown, indicating that the usual Casimir term dominates the source. As $\eta$ increases, the positive memory contribution becomes more relevant near the throat, where the $r^{-7}$ term is strongest, and the density may become positive in the inner region before decreasing and approaching zero from below at larger radii. This behavior illustrates that the memory correction mainly affects the near-throat region, while its influence rapidly weakens as the radial coordinate increases. Therefore, the parameter $\eta$ controls how much the effective source departs from the purely negative Casimir profile close to the wormhole throat.

\subsection{Traversable Lorentzian wormholes}
\label{subsec:lorentzian_wormholes}

We consider the static and spherically symmetric Morris--Thorne line element \cite{MorrisThorne1988}
\begin{equation}
ds^2=-e^{2\Phi(r)}dt^2+\frac{dr^2}{1-\frac{b(r)}{r}}+r^2d\Omega^2,
\label{eq:mt_metric}
\end{equation}
where $b(r)$ is the shape function and $\Phi(r)$ is the redshift function. The coordinate $r$ is the areal radius, so that spheres centered on the throat have area $4\pi r^2$. The function $b(r)$ determines how the spatial geometry departs from the Euclidean one and, in particular, fixes the location and the opening of the throat. The redshift function $\Phi(r)$ controls the gravitational time dilation experienced by static observers. In a traversable wormhole, $\Phi(r)$ must remain finite everywhere, since a divergence of $e^{2\Phi(r)}$ or its vanishing would signal the presence of an event horizon, preventing two-way traversal.

The wormhole throat is defined as the minimum value of the areal radius and is located at $r=r_0$. In the Morris--Thorne parametrization, this condition is encoded in
\begin{equation}
b(r_0)=r_0.
\label{eq:throat_condition_general}
\end{equation}
At this point, the metric coefficient $g_{rr}$ diverges in the coordinate representation, but this is not a physical singularity. It only indicates that the coordinate $r$ reaches a minimum at the throat. The proper radial distance remains well defined provided the geometry satisfies the usual wormhole regularity conditions.

The first important geometric requirement is that the wormhole must flare out at the throat. This means that the spatial slice must open outward instead of closing into a compact surface. In terms of the shape function, the local flare-out condition is
\begin{equation}
b'(r_0)<1.
\label{eq:flareout_local_general}
\end{equation}
Equivalently, in the embedding description, the flare-out behavior is associated with the positivity of the quantity
\begin{equation}
\frac{b(r)-rb'(r)}{b^2(r)}>0
\label{eq:flareout_general}
\end{equation}
near the throat. This condition is purely geometrical and is responsible for the characteristic opening of a traversable wormhole. Through the Einstein equations, it is also the origin of the need for exotic matter, since it forces the violation of the radial null energy condition at the throat.

A second requirement is that the metric preserves its Lorentzian signature outside the throat. For $r>r_0$, one must impose
\begin{equation}
1-\frac{b(r)}{r}>0,
\qquad \text{or equivalently} \qquad b(r)<r.
\label{eq:signature_condition_general}
\end{equation}
This inequality prevents the radial component of the metric from changing sign and guarantees that $r$ remains a spacelike coordinate outside the throat. Thus, while $b(r_0)=r_0$ identifies the throat, the condition $b(r)<r$ ensures that the geometry extends smoothly away from it.

Finally, in order for the wormhole to connect two asymptotically flat regions, the geometry must approach the Minkowski spacetime at large distances. For the shape function, this requires
\begin{equation}
\frac{b(r)}{r}\rightarrow 0
\qquad
\text{as}
\qquad
r\rightarrow \infty.
\label{eq:asymptotic_shape_general}
\end{equation}
For the redshift sector, one also requires $\Phi(r)$ to approach a finite constant at infinity. This constant can be absorbed by a redefinition of the time coordinate, so that the asymptotic region may be normalized in the standard way. Therefore, the complete traversability conditions involve both the spatial sector, controlled by $b(r)$, and the causal sector, controlled by $\Phi(r)$.

These restrictions are imposed before specifying the matter source. They provide the geometric criteria that any density profile must satisfy after being inserted into the Einstein equations. In the present work, the Casimir-memory density determines the shape function, and the above constraints will be used to identify the admissible range of the memory parameter.

\section{Geometric properties}
\label{sec:geometry}

In this section, we derive the geometry generated by the Casimir-memory density profile. We obtain the shape function directly from the Einstein equations, determine the redshift function under a barotropic equation of state, and identify the parameter intervals compatible with traversability. We also discuss curvature diagnostics and embedding diagrams, which provide complementary information about the regularity and spatial structure of the solution.

\subsection{Shape function}
\label{subsec:shape}

The Einstein equations for the metric \eqref{eq:mt_metric}, in geometrized units $G=c=1$, are
\begin{align}
\rho(r)&=\frac{b'(r)}{8\pi r^2},
\label{eq:rho_einstein}\\
p_r(r)&=\frac{1}{8\pi}
\left[
-\frac{b(r)}{r^3}
+2\left(1-\frac{b(r)}{r}\right)\frac{\Phi'(r)}{r}
\right],
\label{eq:pr_einstein}\\
p_t(r)&=\frac{1}{8\pi}\left(1-\frac{b(r)}{r}\right)
\left[
\Phi''+(\Phi')^2
+\frac{b-rb'}{2r(r-b)}\Phi'
+\frac{b-rb'}{2r^2(r-b)}
+\frac{\Phi'}{r}
\right].
\label{eq:pt_einstein}
\end{align}
The pressures are written with lowercase letters to emphasize that they are matter-sector quantities. Once the density profile is specified, Eq.~\eqref{eq:rho_einstein} determines the shape function. The remaining equations then determine the radial and tangential pressures after a redshift function, or an equation of state, is supplied.

Substituting Eq.~\eqref{eq:density_profile} into Eq.~\eqref{eq:rho_einstein}, one obtains
\begin{equation}
b'(r)=8\pi r^2\rho(r)
=-\frac{8\pi\alpha}{r^2}+\frac{8\pi\eta}{r^5}.
\label{eq:bprime}
\end{equation}
Integrating from the throat to an arbitrary radius $r$ and imposing $b(r_0)=r_0$, the shape function becomes
\begin{equation}
b(r)=r_0+8\pi\alpha\left(\frac{1}{r}-\frac{1}{r_0}\right)
-2\pi\eta\left(\frac{1}{r^4}-\frac{1}{r_0^4}\right).
\label{eq:shape_function}
\end{equation}
This expression satisfies the throat condition by construction. It also gives
\begin{equation}
b'(r_0)=
-\frac{8\pi\alpha}{r_0^2}
+\frac{8\pi\eta}{r_0^5}
=
\frac{8\pi}{r_0^5}\left(\eta-\alpha r_0^3\right).
\label{eq:bprime_throat}
\end{equation}
The sign of $b'(r_0)$ is controlled by the balance between the ordinary Casimir term and the memory term. Therefore, the gravitational memory does not merely change the magnitude of the source; it directly modifies the local geometry at the throat.

It is also useful to note that Eq.~\eqref{eq:shape_function} is asymptotically compatible with flatness, since $b(r)/r\to0$ as $r\to\infty$. Nevertheless, the condition $b(r)<r$ for all $r>r_0$ must be checked globally for each choice of parameters. This will be addressed in Sec.~\ref{subsec:geometric_constraints}.

\subsection{Redshift function and barotropic equation of state}
\label{subsec:redshift}

The geometry is not completely determined by the density profile alone. After $b(r)$ is obtained, one still needs a closure relation involving the pressure sector or the redshift function. We adopt a barotropic equation of state for the radial pressure,
\begin{equation}
p_r(r)=\omega \rho(r),
\label{eq:eos}
\end{equation}
where $\omega$ is taken as a constant. Combining Eqs.~\eqref{eq:rho_einstein}, \eqref{eq:pr_einstein}, and \eqref{eq:eos}, one obtains
\begin{equation}
\Phi'(r)=
\frac{b(r)+\omega r b'(r)}{2r\left[r-b(r)\right]}.
\label{eq:phiprime_general}
\end{equation}
This expression shows that the redshift function is strongly constrained by the throat behavior. Since $r-b(r)$ vanishes at $r=r_0$, the numerator must also vanish there in order to avoid a logarithmic divergence in $\Phi(r)$. Thus, $b(r_0)+\omega r_0 b'(r_0)=0$, which gives
\begin{equation}
\omega=-\frac{1}{b'(r_0)}
=
\frac{r_0^5}{8\pi\left(\alpha r_0^3-\eta\right)}.
\label{eq:omega_regular}
\end{equation}
The value $\eta=\alpha r_0^3$ is therefore singular in this barotropic description, because it implies $b'(r_0)=0$ and would require $|\omega|\to\infty$.

For later use, it is convenient to rewrite the shape function as
\begin{equation}
b(r)=b_0+\frac{\mathcal{A}}{r}-\frac{\mathcal{B}}{r^4},
\qquad
\mathcal{A}=8\pi\alpha,
\qquad
\mathcal{B}=2\pi\eta,
\label{eq:shape_AB}
\end{equation}
where the throat condition fixes $b_0=r_0-\mathcal{A}/r_0+\mathcal{B}/r_0^4$. In this notation,
\begin{equation}
\Phi'(r)=
\frac{b_0 r^4+\mathcal{A}(1-\omega)r^3+\mathcal{B}(4\omega-1)}
{2r\left(r^5-b_0r^4-\mathcal{A}r^3+\mathcal{B}\right)}.
\label{eq:phiprime_explicit}
\end{equation}
The denominator contains the polynomial
\begin{equation}
P_5(r)=r^5-b_0r^4-\mathcal{A}r^3+\mathcal{B}
=r^4\left[r-b(r)\right],
\label{eq:p5_polynomial}
\end{equation}
whose roots will be denoted by $s_i$. Since the throat condition imposes $b(r_0)=r_0$, one of these roots is always $s_i=r_0$. The integrated redshift function can then be written as
\begin{equation}
\Phi(r)=\Phi_0+\frac{1}{2}(4\omega-1)\ln r
-\frac{1}{2}\sum_{i=1}^{5}
\frac{3\mathcal{A}\omega+4b_0\omega s_i+(1-4\omega)s_i^2}
{3\mathcal{A}+4b_0s_i-5s_i^2}
\ln(r-s_i).
\label{eq:redshift_integrated}
\end{equation}
For complex roots, the corresponding logarithmic terms must be combined in conjugate pairs so that the physical redshift function remains real. The arbitrary constants associated with the logarithmic branches can be absorbed into $\Phi_0$. The apparent logarithmic divergence associated with the throat root is removed when the regularity condition \eqref{eq:omega_regular} is imposed. Thus, Eq.~\eqref{eq:redshift_integrated} gives the explicit redshift function compatible with the barotropic source and with the absence of horizons at the throat.

\subsection{Geometric constraints}
\label{subsec:geometric_constraints}

The local flare-out condition at the throat requires $b'(r_0)<1$. Using Eq.~\eqref{eq:bprime_throat}, this gives
\begin{equation}
\eta<\alpha r_0^3+\frac{r_0^5}{8\pi}.
\label{eq:eta_flareout}
\end{equation}
Together with the basic physical ranges $\alpha>0$ and $\eta\ge0$, the local geometric domain is
\begin{equation}
\alpha>0,\qquad
0\leq\eta<\alpha r_0^3+\frac{r_0^5}{8\pi}.
\label{eq:eta_total_range}
\end{equation}
Inside this domain, two regimes must be distinguished. If $0\leq\eta<\alpha r_0^3$, then $b'(r_0)<0$ and the regularity condition \eqref{eq:omega_regular} gives $\omega>0$. This is the Casimir-dominated regime. If
\begin{equation}
\alpha r_0^3<\eta<\alpha r_0^3+\frac{r_0^5}{8\pi},
\label{eq:eta_phantom}
\end{equation}
then $0<b'(r_0)<1$ and Eq.~\eqref{eq:omega_regular} gives $\omega<-1$. This is a phantom-like regime induced by a sufficiently strong memory contribution.

The critical value $\eta=\alpha r_0^3$ separates these two sectors. At this point the density at the throat vanishes and the barotropic description becomes singular, so this value must be excluded from the constant-$\omega$ model. It is important to emphasize that the interval allowed by the flare-out condition is a geometric domain of the effective wormhole solution. It is broader than the strictly perturbative regime associated with the original Casimir-memory calculation. Therefore, values satisfying $\eta/(\alpha r_0^3)\ll1$ should be regarded as the conservative sector directly connected with a small memory correction, whereas values closer to or above $\alpha r_0^3$ explore an effective phenomenological extension of the same density profile.

The remaining geometric conditions, namely $b(r)<r$ for $r>r_0$ and $b(r)/r\to0$, must be examined globally. The second one is automatically satisfied by Eq.~\eqref{eq:shape_function}. The first one depends on the interplay among $\alpha$, $\eta$, and $r_0$, and will be illustrated numerically.

For the numerical analysis of the geometric constraints, we fix the ordinary Casimir parameter at its electromagnetic value, $\alpha=\pi^2/720$, and set $r_0=1$ in geometrized units. With this choice, the critical value separating the Casimir-dominated and memory-dominated sectors is $\eta_c=\alpha r_0^3\simeq0.01371$, while the upper value allowed by the local flare-out condition is $\eta_{\max}=\alpha r_0^3+r_0^5/(8\pi)\simeq0.05350$. We use four representative values, $\eta=0.00274$, $0.01234$, $0.02365$, and $0.04355$, in order to sample different regions of the admissible geometric interval. The first two values remain in the Casimir-dominated sector, with the second one lying close to the transition scale, while the last two values probe the extended effective sector in which the memory correction has a substantial influence on the near-throat geometry.

\begin{figure}[!htp]
\centering
\includegraphics[width=0.49\textwidth]{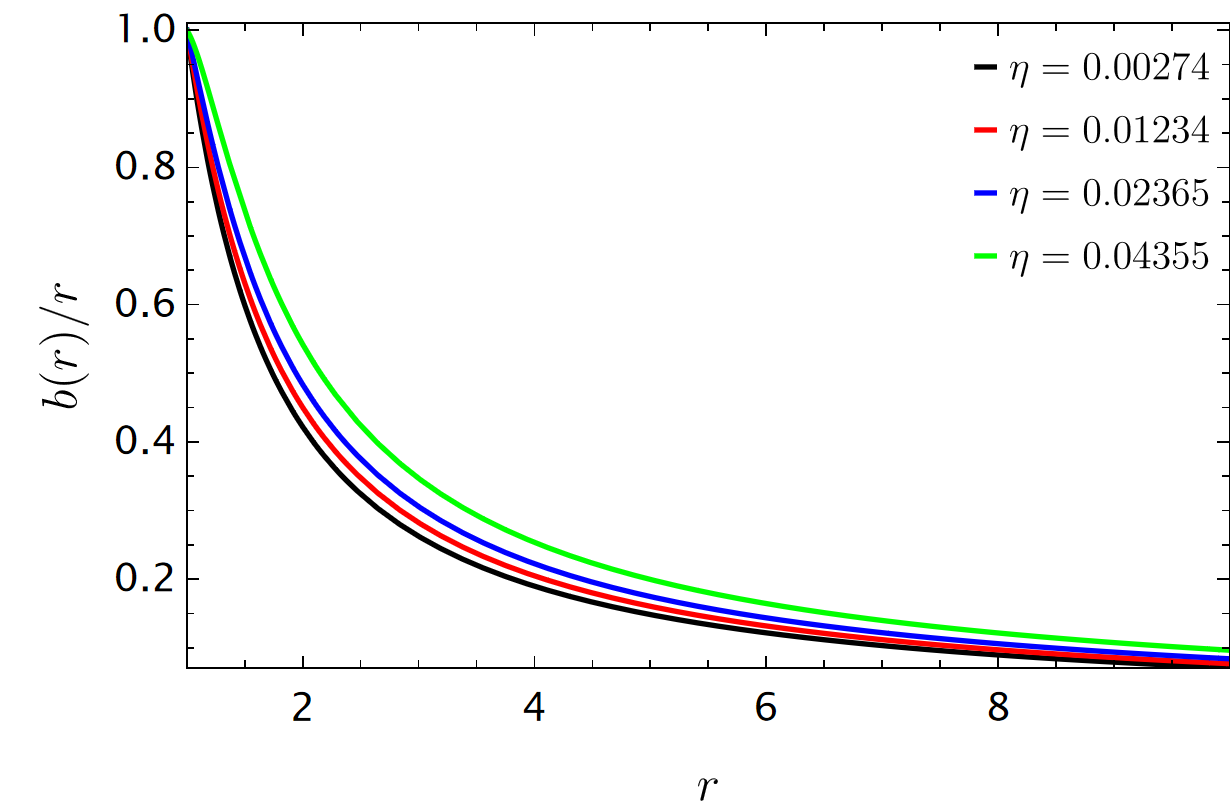}
\includegraphics[width=0.49\linewidth]{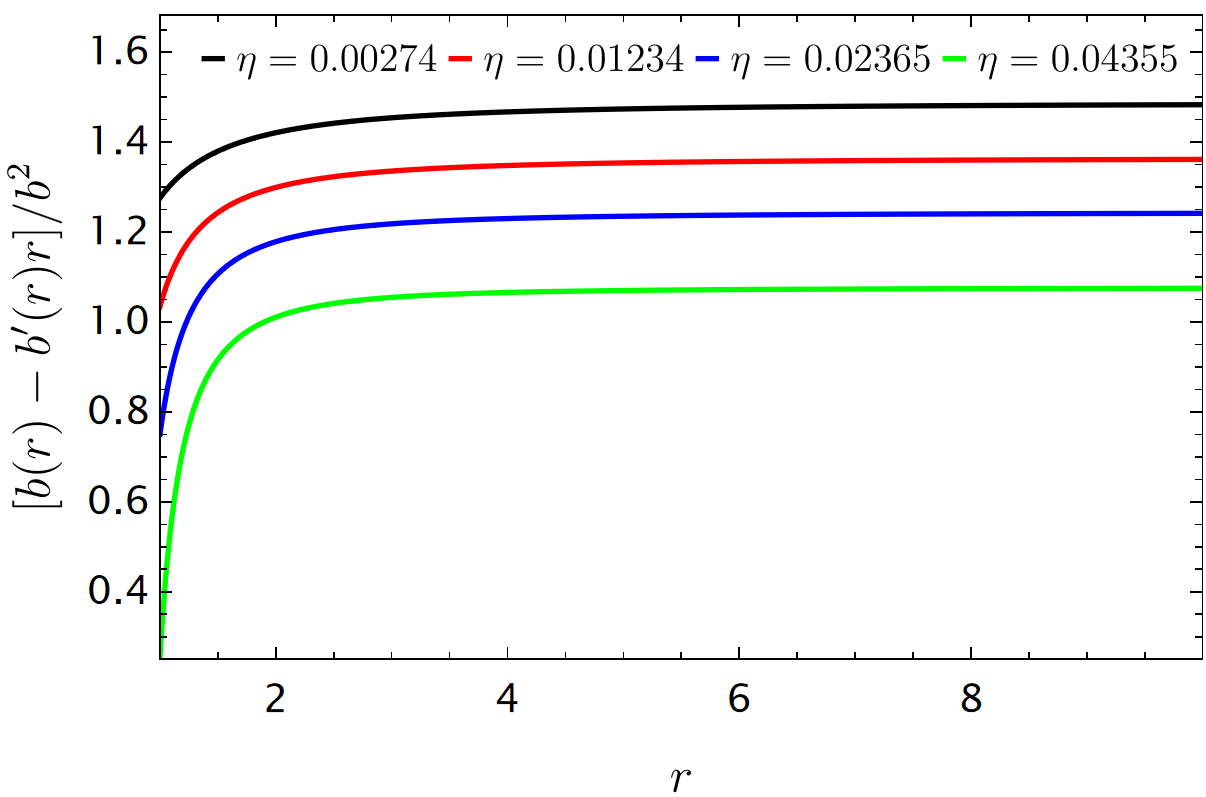}
\caption{Geometric behavior of the Casimir-memory wormhole for different values of the memory parameter $\eta$, with $\alpha=\pi^2/720$ and $r_0=1$ fixed. In the left panel, we show the ratio $b(r)/r$ as a function of the radial coordinate $r$, illustrating the asymptotic behavior of the shape function. In the right panel, we present the flare-out quantity $[b(r)-rb'(r)]/b^2(r)$. The curves correspond to $\eta=0.00274$, $0.01234$, $0.02365$, and $0.04355$.}
\label{fig:geometric_constraints}
\end{figure}

Figure~\ref{fig:geometric_constraints} displays the main geometric requirements associated with the shape function generated by the Casimir-memory source. In the left panel, all curves start from $b(r_0)/r_0=1$, as required by the throat condition, and decrease as the radial coordinate increases. The ratio $b(r)/r$ tends to zero for any admissible values of $\alpha$ and $\eta$, since the shape function approaches a finite asymptotic value while the denominator grows linearly with $r$. This guarantees that the spacetime is asymptotically flat independently of the particular strength of the memory correction. Increasing $\eta$ shifts the curves upward in the intermediate region, showing that the positive memory contribution increases the effective shape function outside the throat, but without spoiling the condition $b(r)<r$ for $r>r_0$. The right panel shows that the flare-out quantity remains positive for the chosen values of $\eta$, confirming that the spatial geometry opens outward from the throat. The decrease of this quantity as $\eta$ increases indicates that the memory correction softens the flare-out behavior, but the geometry still remains compatible with a traversable wormhole throughout the plotted domain.

\subsection{Curvatures}
\label{subsec:curvatures}

Curvature scalars provide a direct diagnostic of the regularity of the geometry. In particular, the Ricci scalar may be written in terms of the matter variables as
\begin{equation}
R(r)=8\pi\left[\rho(r)-p_r(r)-2p_t(r)\right].
\label{eq:ricci_matter}
\end{equation}
Equivalently, for the Morris--Thorne metric \eqref{eq:mt_metric},
\begin{equation}
R(r)=
\frac{
2b'
+\left(rb'-b\right)\Phi'
-4(r-b)\Phi'
-2r(r-b)\left[\Phi''+(\Phi')^2\right]
}{r^2}.
\label{eq:ricci_geometry}
\end{equation}
Using the explicit derivative \eqref{eq:bprime}, this becomes
\begin{align}
R(r)=&
-\frac{16\pi\alpha}{r^4}
+\frac{16\pi\eta}{r^7}
+\frac{\left[rb'(r)-b(r)-4(r-b(r))\right]\Phi'(r)}{r^2}
\nonumber\\
&-2\frac{r-b(r)}{r}\left[\Phi''(r)+(\Phi'(r))^2\right],
\label{eq:ricci_explicit}
\end{align}
where $b(r)$ is given by Eq.~\eqref{eq:shape_function}. This expression makes explicit how the Ricci scalar receives a direct contribution from the Casimir-memory density and additional contributions from the redshift sector. In the zero-tidal-force limit, $\Phi'=0$, one recovers $R(r)=2b'(r)/r^2=-16\pi\alpha/r^4+16\pi\eta/r^7$.

At the throat, assuming a finite redshift derivative, Eq.~\eqref{eq:ricci_geometry} reduces to
\begin{equation}
R(r_0)=
\frac{2b'(r_0)+r_0\left[b'(r_0)-1\right]\Phi'(r_0)}{r_0^2}.
\label{eq:ricci_throat}
\end{equation}
Thus, the throat curvature is controlled by both the flare-out combination $b'(r_0)-1$ and the redshift gradient. The memory parameter enters through $b'(r_0)$ and can either soften or intensify the local curvature depending on its value relative to $\alpha r_0^3$.

\begin{figure}[!htp]
\centering
\includegraphics[width=0.6\textwidth]{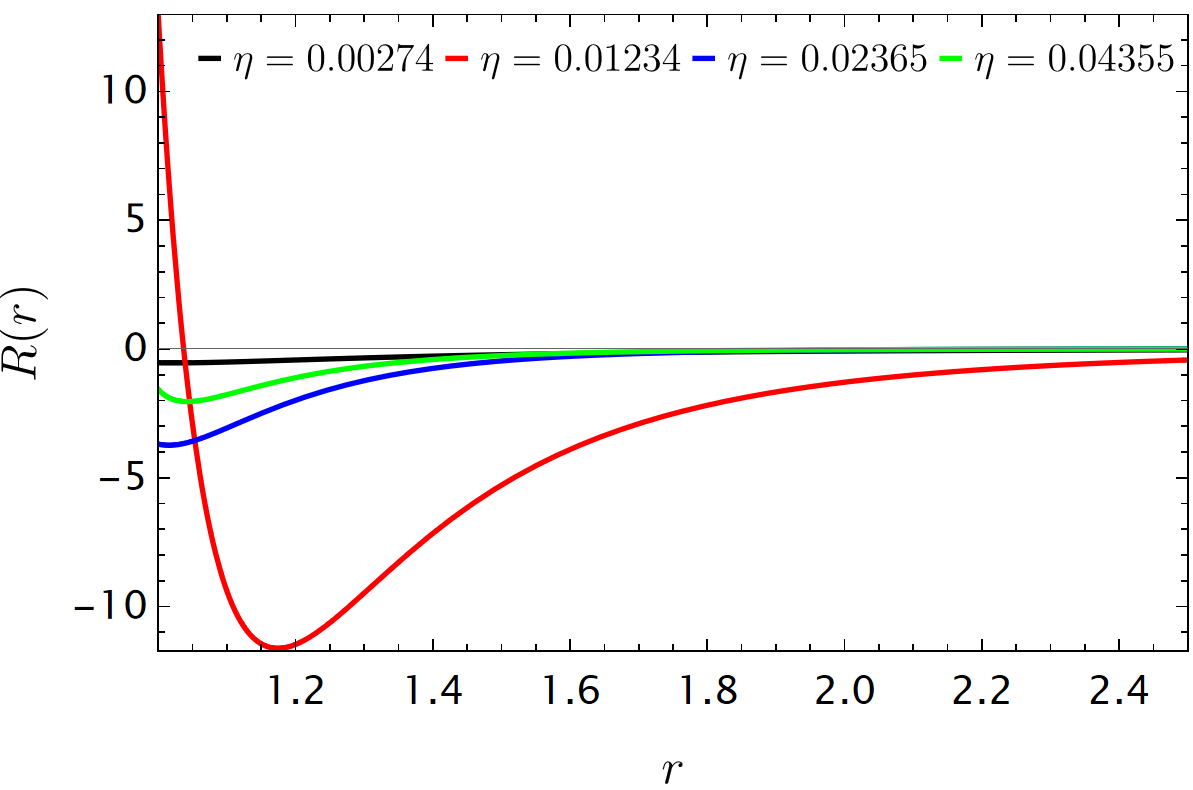}
\caption{Ricci scalar $R(r)$ for the Casimir-memory wormhole as a function of the radial coordinate $r$, for different values of the memory parameter $\eta$, with $\alpha=\pi^2/720$ and $r_0=1$ fixed. The curves correspond to $\eta=0.00274$, $0.01234$, $0.02365$, and $0.04355$. The behavior near the throat shows that the curvature is strongly affected by the memory contribution, especially for values of $\eta$ close to the transition scale $\eta_c=\alpha r_0^3$, while all curves tend to zero as $r$ increases, consistently with the asymptotically flat character of the geometry.}
\label{fig:ricci_scalar}
\end{figure}

Figure~\ref{fig:ricci_scalar} shows that the curvature is mainly concentrated near the throat and rapidly decreases as the radial coordinate increases. For all representative values of $\eta$, the Ricci scalar tends to zero at large $r$, consistently with the asymptotically flat behavior already indicated by $b(r)/r\to0$. The most pronounced variation occurs for $\eta=0.01234$, which lies close to the transition scale $\eta_c=\alpha r_0^3$. In this case, the redshift sector becomes more sensitive to the near-throat geometry, producing a sharper curvature profile. Away from this transition region, the memory parameter changes the intensity of the curvature in a smoother way: the curves for $\eta=0.02365$ and $\eta=0.04355$ remain less steep and approach the asymptotic regime more regularly. Therefore, the Ricci scalar confirms that the gravitational-memory correction affects primarily the strong-field region around the throat, while the spacetime becomes progressively weakly curved as one moves toward the asymptotic domain.

\subsection{Embedding diagram}
\label{subsec:embedding}

The spatial geometry of the wormhole can be visualized through an embedding diagram. We consider a constant-time equatorial slice, $t={\rm const.}$ and $\theta=\pi/2$, for which the line element becomes
\begin{equation}
ds^2=\frac{dr^2}{1-\frac{b(r)}{r}}+r^2d\phi^2.
\label{eq:equatorial_slice}
\end{equation}
This two-dimensional surface can be embedded in a three-dimensional Euclidean space with cylindrical metric
\begin{equation}
ds^2=dz^2+dr^2+r^2d\phi^2.
\label{eq:euclidean_embedding}
\end{equation}
Assuming axial symmetry, the embedded surface is described by $z=z(r)$, so that
\begin{equation}
ds^2=\left[1+\left(\frac{dz}{dr}\right)^2\right]dr^2+r^2d\phi^2.
\end{equation}
Comparison with Eq.~\eqref{eq:equatorial_slice} gives
\begin{equation}
\frac{dz}{dr}=
\pm\sqrt{\frac{b(r)}{r-b(r)}}.
\label{eq:embedding_equation}
\end{equation}
The divergence of $dz/dr$ at $r=r_0$ is not a curvature singularity; it is the usual geometrical indication that the embedded surface becomes vertical at the throat. The flare-out condition guarantees that the embedded surface opens outward rather than closing.

The profile $z(r)$ is then obtained from
\begin{equation}
z(r)=\pm\int_{r_0}^{r}
\sqrt{\frac{b(x)}{x-b(x)}}\,dx.
\label{eq:z_integral}
\end{equation}
For the shape function \eqref{eq:shape_function}, this integral is most conveniently evaluated numerically. Revolving the resulting curve around the vertical axis produces the three-dimensional embedding surface.

\begin{figure}[!htp]
\centering
\includegraphics[width=0.55\textwidth]{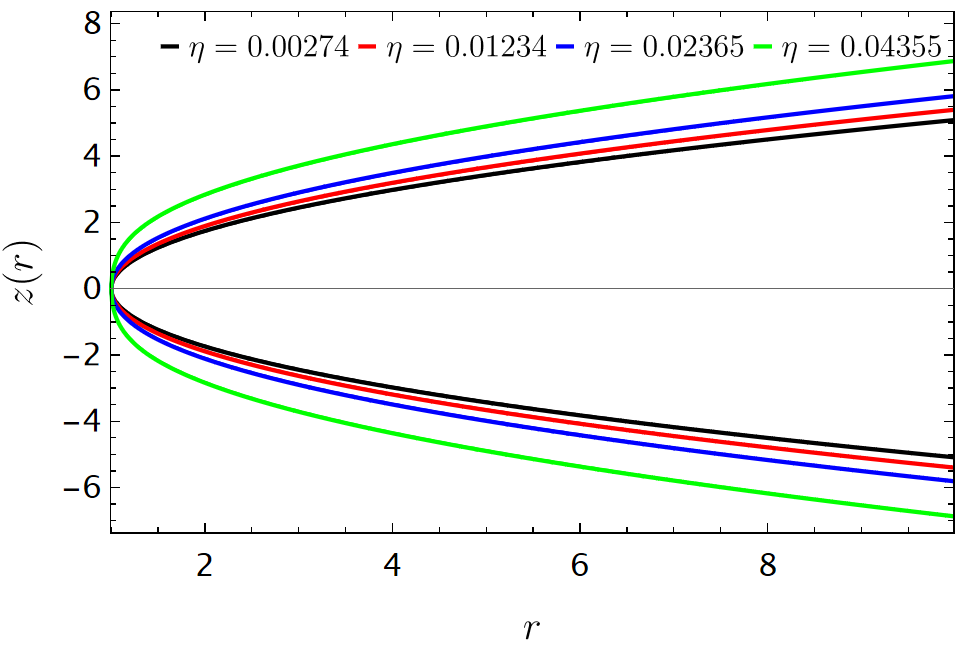}
\includegraphics[width=0.4\textwidth]{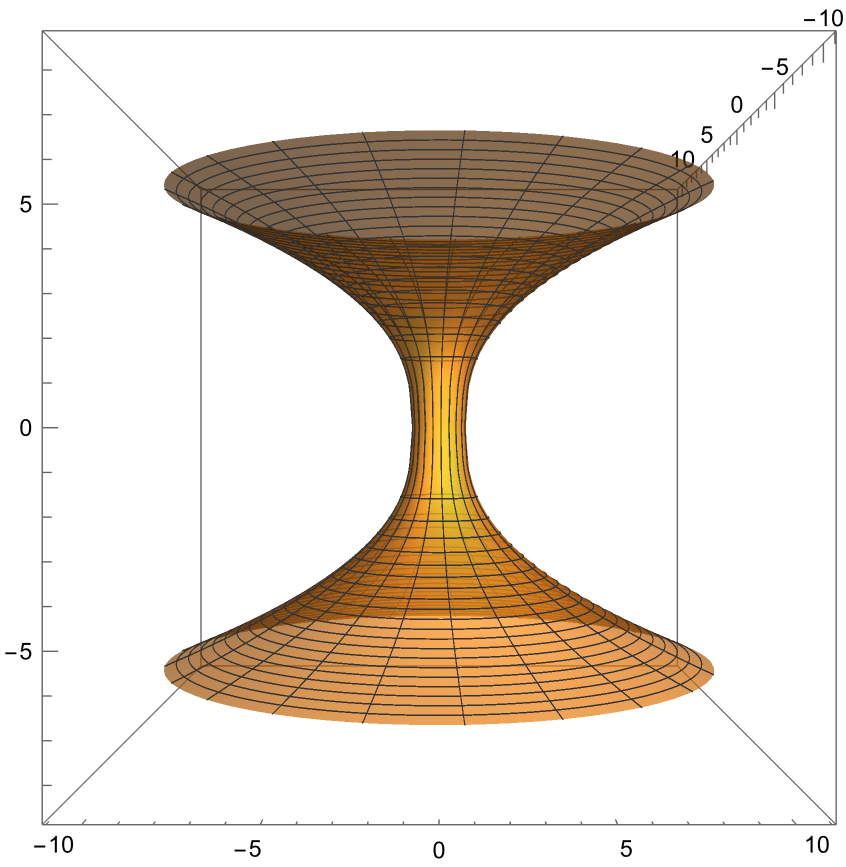}
\caption{Embedding structure of the Casimir-memory wormhole geometry for different values of the memory parameter $\eta$, with $\alpha=\pi^2/720$ and $r_0=1$ fixed. In the left panel, we show the embedding function $z(r)$ as a function of the radial coordinate $r$ for $\eta=0.00274$, $0.01234$, $0.02365$, and $0.04355$. In the right panel, we present the corresponding three-dimensional embedded surface for $\eta=0.04355$, illustrating the spatial geometry generated by the Casimir-memory shape function.}
\label{fig:embedding}
\end{figure}

Figure~\ref{fig:embedding} illustrates the spatial embedding associated with the Casimir-memory shape function, with $r_0=1$ fixed in all cases. The two-dimensional profiles show the expected vertical behavior at the throat and then open toward the asymptotic region, confirming the wormhole character of the spatial slice. The corrected profiles also show a monotonic dependence on the memory parameter: for a fixed radial coordinate $r>r_0$, the magnitude of the embedding function increases as $\eta$ increases. This behavior follows directly from the shape function, since the memory contribution increases $b(r)$ outside the throat and consequently enhances the embedding integrand $[b(r)/(r-b(r))]^{1/2}$. Thus, although the throat radius remains fixed, increasing $\eta$ produces a progressively wider spatial embedding away from the throat. For the largest value, $\eta=0.04355$, this effect is most pronounced, as shown by both the outermost two-dimensional profile and the corresponding three-dimensional surface. The right panel provides a geometric visualization of this behavior: the memory correction widens the spatial geometry outside the throat while preserving the smooth connection between the two asymptotic regions.

\section{Matter distribution, energy conditions and equilibrium analysis}
\label{sec:matter}

Having established the geometric structure of the solution, we now analyze the matter content required to sustain it. We first write the energy conditions in terms of the effective density and pressures, with special attention to the throat. We then examine the equilibrium of the anisotropic source using the Tolman--Oppenheimer--Volkoff equation.

\subsection{Energy conditions}
\label{subsec:energy_conditions}

For an anisotropic matter distribution,
\begin{equation}
T^{\mu}{}_{\nu}={\rm diag}\left[-\rho(r),p_r(r),p_t(r),p_t(r)\right].
\label{eq:stress_tensor}
\end{equation}
The main energy conditions are expressed through the combinations
\begin{align}
{\rm NEC}_r&:\quad \rho+p_r\ge0,\\
{\rm NEC}_t&:\quad \rho+p_t\ge0,\\
{\rm WEC}&:\quad \rho\ge0,\quad \rho+p_r\ge0,\quad \rho+p_t\ge0,\\
{\rm SEC}&:\quad \rho+p_r+2p_t\ge0,\quad \rho+p_r\ge0,\quad \rho+p_t\ge0,\\
{\rm DEC}&:\quad \rho-|p_r|\ge0,\quad \rho-|p_t|\ge0.
\end{align}
For a traversable wormhole, the radial NEC must be violated at the throat. This follows directly from the Einstein equations and the flare-out condition.

The density profile is the one given in Eq.~\eqref{eq:density_profile}, while the shape function and the redshift derivative are those obtained in Eqs.~\eqref{eq:shape_AB} and \eqref{eq:phiprime_explicit}, respectively. Since the radial pressure is fixed by the barotropic relation, the regularity condition \eqref{eq:omega_regular} allows us to write it directly as \begin{equation} p_r(r)= \frac{r_0^5}{8\pi\left(\alpha r_0^3-\eta\right)} \left( -\frac{\alpha}{r^4}+\frac{\eta}{r^7} \right). \label{eq:pr_explicit} \end{equation} Therefore, the radial sector is completely determined by the Casimir parameter $\alpha$, the memory parameter $\eta$, and the throat radius $r_0$. The tangential pressure, on the other hand, is obtained from the angular component of the Einstein equations after substituting the explicit shape function and the redshift derivative. Using the notation already introduced in Eq.~\eqref{eq:shape_AB}, it reads \begin{align} p_t(r)=\frac{1}{8\pi} \left( 1-\frac{b_0}{r}-\frac{\mathcal{A}}{r^2} +\frac{\mathcal{B}}{r^5} \right) \Bigg\{ & \frac{d}{dr} \left[ \frac{ b_0 r^4+\mathcal{A} \left( 1-\frac{r_0^5}{8\pi(\alpha r_0^3-\eta)} \right)r^3 +\mathcal{B} \left( \frac{r_0^5}{2\pi(\alpha r_0^3-\eta)}-1 \right) } { 2r\left(r^5-b_0r^4-\mathcal{A}r^3+\mathcal{B}\right) } \right] \nonumber\\ &+ \left[ \frac{ b_0 r^4+\mathcal{A} \left( 1-\frac{r_0^5}{8\pi(\alpha r_0^3-\eta)} \right)r^3 +\mathcal{B} \left( \frac{r_0^5}{2\pi(\alpha r_0^3-\eta)}-1 \right) } { 2r\left(r^5-b_0r^4-\mathcal{A}r^3+\mathcal{B}\right) } \right]^2 \nonumber\\ &+ \frac{ b_0+\frac{2\mathcal{A}}{r}-\frac{5\mathcal{B}}{r^4} }{ 2r^2 \left( 1-\frac{b_0}{r}-\frac{\mathcal{A}}{r^2} +\frac{\mathcal{B}}{r^5} \right) } \left[ \frac{ b_0 r^4+\mathcal{A} \left( 1-\frac{r_0^5}{8\pi(\alpha r_0^3-\eta)} \right)r^3 +\mathcal{B} \left( \frac{r_0^5}{2\pi(\alpha r_0^3-\eta)}-1 \right) } { 2r\left(r^5-b_0r^4-\mathcal{A}r^3+\mathcal{B}\right) } \right] \nonumber\\ &+ \frac{ b_0+\frac{2\mathcal{A}}{r}-\frac{5\mathcal{B}}{r^4} }{ 2r^3 \left( 1-\frac{b_0}{r}-\frac{\mathcal{A}}{r^2} +\frac{\mathcal{B}}{r^5} \right) } \nonumber\\ &+ \frac{1}{r} \left[ \frac{ b_0 r^4+\mathcal{A} \left( 1-\frac{r_0^5}{8\pi(\alpha r_0^3-\eta)} \right)r^3 +\mathcal{B} \left( \frac{r_0^5}{2\pi(\alpha r_0^3-\eta)}-1 \right) } { 2r\left(r^5-b_0r^4-\mathcal{A}r^3+\mathcal{B}\right) } \right] \Bigg\}, \label{eq:pt_explicit} \end{align} 
where $\mathcal{A} = 8\pi\alpha$ and $\mathcal{B} = 2\pi\eta$. This expression makes explicit that the tangential pressure is not imposed by the equation of state, but reconstructed from the field equations once the Casimir-memory density and the regular redshift sector have been fixed.

At the throat, these expressions reduce to
\begin{equation}
\rho(r_0)=
-\frac{\alpha}{r_0^4}
+\frac{\eta}{r_0^7}
=
\frac{b'(r_0)}{8\pi r_0^2},
\label{eq:rho_throat}
\end{equation}
\begin{equation}
p_r(r_0)=
-\frac{1}{8\pi r_0^2},
\label{eq:pr_throat}
\end{equation}
and
\begin{equation}
p_t(r_0)=
\frac{1-b'(r_0)}{16\pi r_0^2}
\left[1+r_0\Phi'(r_0)\right],
\label{eq:pt_throat}
\end{equation}
provided $\Phi'(r_0)$ is finite.

The radial NEC for arbitrary $r$ is
\begin{equation}
\rho+p_r=
\frac{1}{8\pi}
\left[
\frac{b'}{r^2}
-\frac{b}{r^3}
+2\left(1-\frac{b}{r}\right)\frac{\Phi'}{r}
\right],
\label{eq:nec_r_general}
\end{equation}
and at the throat it becomes
\begin{equation}
\left.\left(\rho+p_r\right)\right|_{r=r_0}
=
\frac{b'(r_0)-1}{8\pi r_0^2}.
\label{eq:nec_r_throat}
\end{equation}
Therefore, the flare-out condition $b'(r_0)<1$ implies $\rho(r_0)+p_r(r_0)<0$. The radial NEC is necessarily violated at the throat, independently of the detailed form of the redshift function.

The tangential NEC at the throat is
\begin{equation}
\left.\left(\rho+p_t\right)\right|_{r=r_0}
=
\frac{
2b'(r_0)+\left[1-b'(r_0)\right]\left[1+r_0\Phi'(r_0)\right]
}{16\pi r_0^2}.
\label{eq:nec_t_throat}
\end{equation}
Unlike the radial NEC, this condition depends on the redshift gradient. This shows that the exotic character required to open the throat is primarily radial. The tangential sector can remain non-exotic for suitable choices of $\Phi'(r_0)$.

For the barotropic model, the regularity condition \eqref{eq:omega_regular} also gives $p_r(r_0)=\omega\rho(r_0)=-1/(8\pi r_0^2)$, consistently with Eq.~\eqref{eq:pr_throat}. Therefore, the barotropic parameter is not free once the throat is required to be regular. It is fixed by the derivative of the shape function at the throat and, consequently, by the ratio between the ordinary Casimir term and the memory correction.

\begin{figure}[!htp]
\centering
\includegraphics[width=0.49\textwidth]{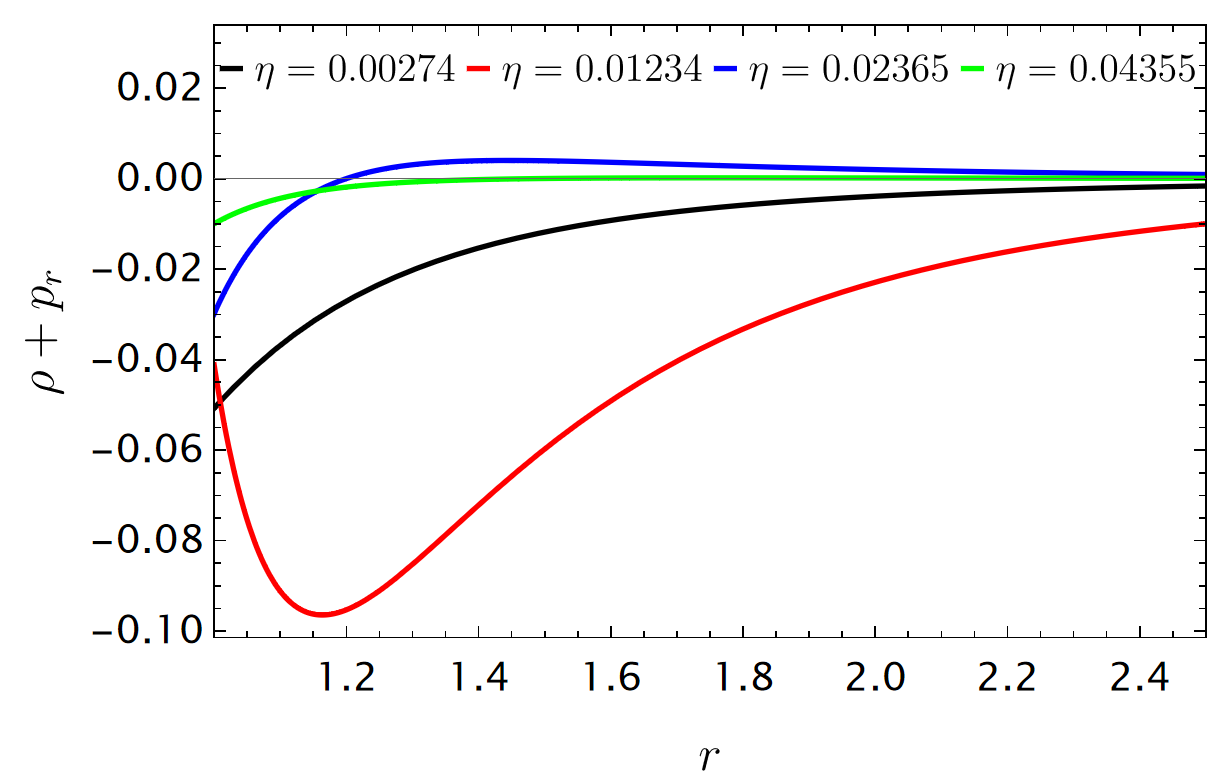}
\includegraphics[width=0.49\linewidth]{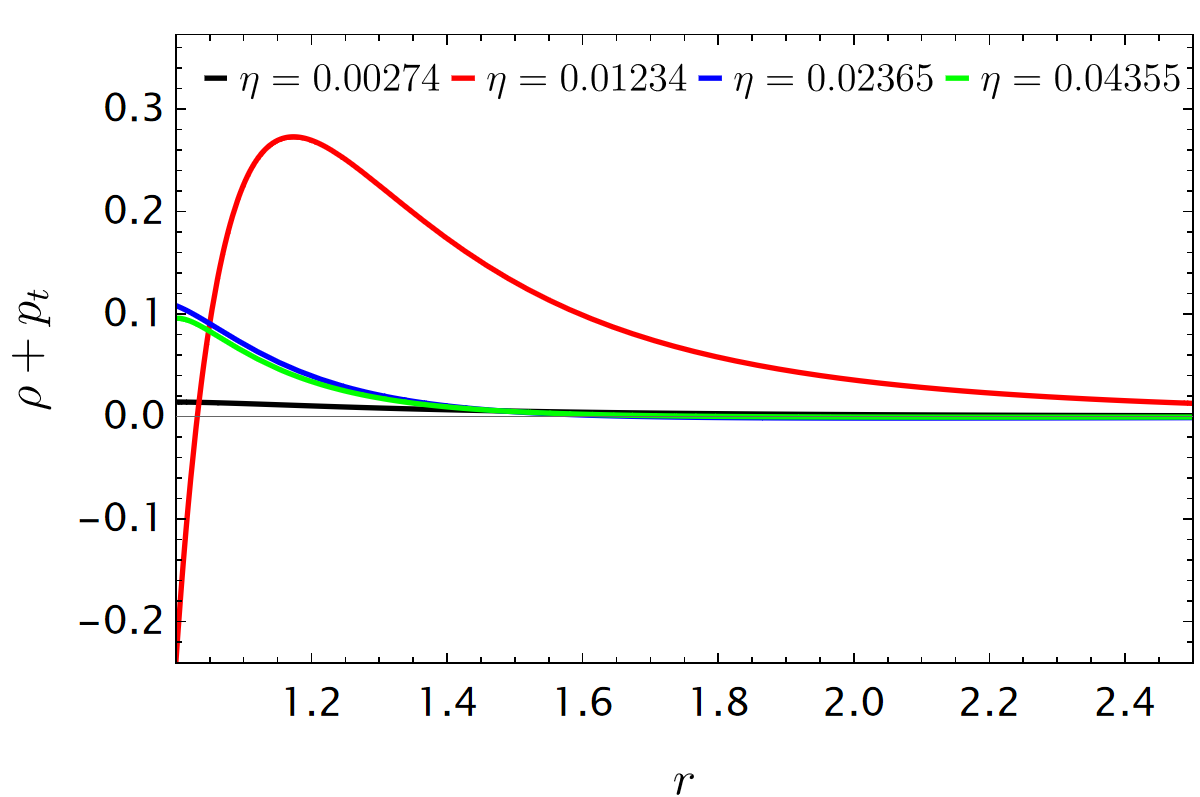}
\includegraphics[width=0.55\linewidth]{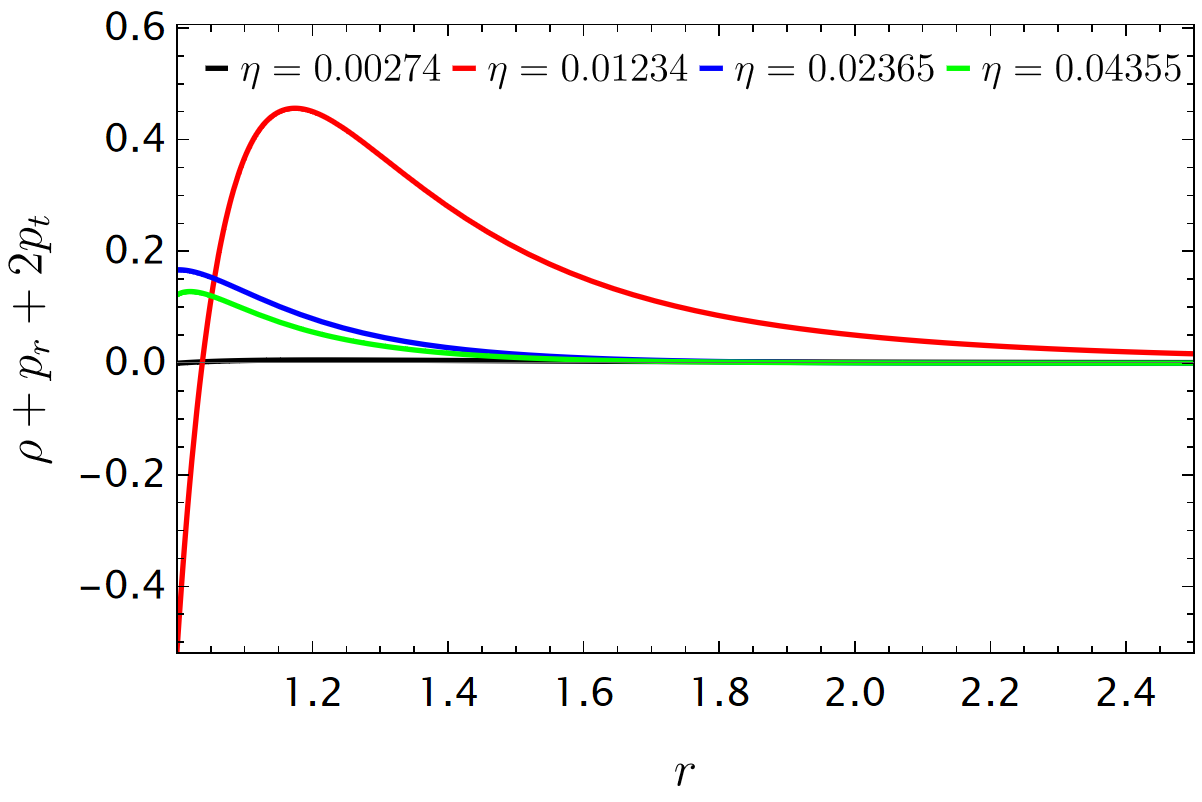}
\caption{Behavior of the energy-condition combinations for the Casimir-memory wormhole, with $\alpha=\pi^2/720$ and $r_0=1$ fixed. The top-left panel shows the radial NEC, $\rho+p_r$, the top-right panel shows the tangential NEC, $\rho+p_t$, and the bottom panel shows the strong energy condition combination, $\rho+p_r+2p_t$. The curves correspond to $\eta=0.00274$, $0.01234$, $0.02365$, and $0.04355$. In all panels, the barotropic parameter $\omega$ is fixed for each value of $\eta$ by the throat-regularity condition \eqref{eq:omega_regular}.}
\label{fig:energy_conditions}
\end{figure}

Figure~\ref{fig:energy_conditions} shows how the main energy-condition combinations respond to the gravitational-memory contribution. The radial NEC, displayed in the top-left panel, is violated at the throat for all representative values of $\eta$, as required by the flare-out condition. For $r_0=1$, the throat values are $(\rho+p_r)_{r_0}\simeq -0.05076$, $-0.04116$, $-0.02985$, and $-0.00995$ for $\eta=0.00274$, $0.01234$, $0.02365$, and $0.04355$, respectively. Thus, increasing $\eta$ weakens the magnitude of the radial NEC violation at the throat, although it does not remove it within the admissible wormhole sector. This is consistent with the physical role of the positive memory term: it partially compensates the negative Casimir contribution, but the geometry still requires a negative radial-energy combination in order to keep the throat open. The behavior away from the throat is not monotonic in a simple way, especially for $\eta=0.01234$, whose curve develops a deeper negative region before approaching zero. This occurs because this value lies close to the transition scale $\eta_c=\alpha r_0^3$, making the redshift sector more sensitive to the throat-regularity condition.

The tangential NEC and the SEC, shown in the top-right and bottom panels, have a more model-dependent behavior because they involve the tangential pressure reconstructed from the angular Einstein equation. At the throat, the tangential NEC gives $(\rho+p_t)_{r_0}\simeq 0.01416$, $-0.24041$, $0.10828$, and $0.09577$, while the SEC combination gives $(\rho+p_r+2p_t)_{r_0}\simeq -0.00050$, $-0.51924$, $0.16684$, and $0.12191$ for the same sequence of $\eta$ values. Therefore, unlike the radial NEC, the tangential NEC and the SEC are not necessarily violated at the throat for all memory strengths. The strongest violation again occurs for $\eta=0.01234$, reflecting the enhanced sensitivity near the critical value of the memory parameter. For the larger values $\eta=0.02365$ and $\eta=0.04355$, both the tangential NEC and the SEC are positive at the throat and decay smoothly toward zero as $r$ increases. This indicates that the exotic character of the source is mainly concentrated in the radial sector, while the tangential sector can become effectively non-exotic depending on the balance between the Casimir term, the memory correction, and the redshift profile.

The equilibrium of an anisotropic static configuration is governed by the conservation equation $\nabla_\mu T^{\mu}{}_{\nu}=0$. For the metric \eqref{eq:mt_metric}, the radial component gives the Tolman--Oppenheimer--Volkoff equation
\begin{equation}
p_r'(r)+\left[\rho(r)+p_r(r)\right]\Phi'(r)
+\frac{2}{r}\left[p_r(r)-p_t(r)\right]=0.
\label{eq:tov}
\end{equation}
This equation can be written as a balance of forces,
\begin{equation}
\mathcal{F}_g+\mathcal{F}_h+\mathcal{F}_a=0,
\label{eq:force_balance}
\end{equation}
where
\begin{equation}
\mathcal{F}_g=-\left(\rho+p_r\right)\Phi',
\qquad
\mathcal{F}_h=-p_r',
\qquad
\mathcal{F}_a=\frac{2}{r}\left(p_t-p_r\right).
\label{eq:forces}
\end{equation}
Here $\mathcal{F}_g$ is the gravitational contribution, $\mathcal{F}_h$ is the hydrostatic force, and $\mathcal{F}_a$ is the anisotropic force. No additional force associated with a running gravitational coupling appears in the present model, because the gravitational memory is encoded in the matter density rather than in a scale-dependent Newton constant.

This distinction is important. The memory correction changes the equilibrium indirectly: it modifies $\rho(r)$, then $b(r)$, then the pressure profiles and the redshift function. The TOV balance therefore provides a diagnostic of how the remembered gravitational perturbation reshapes the internal stress distribution required to sustain the wormhole.

\begin{figure}[!htp]
\centering
\includegraphics[width=0.49\textwidth]{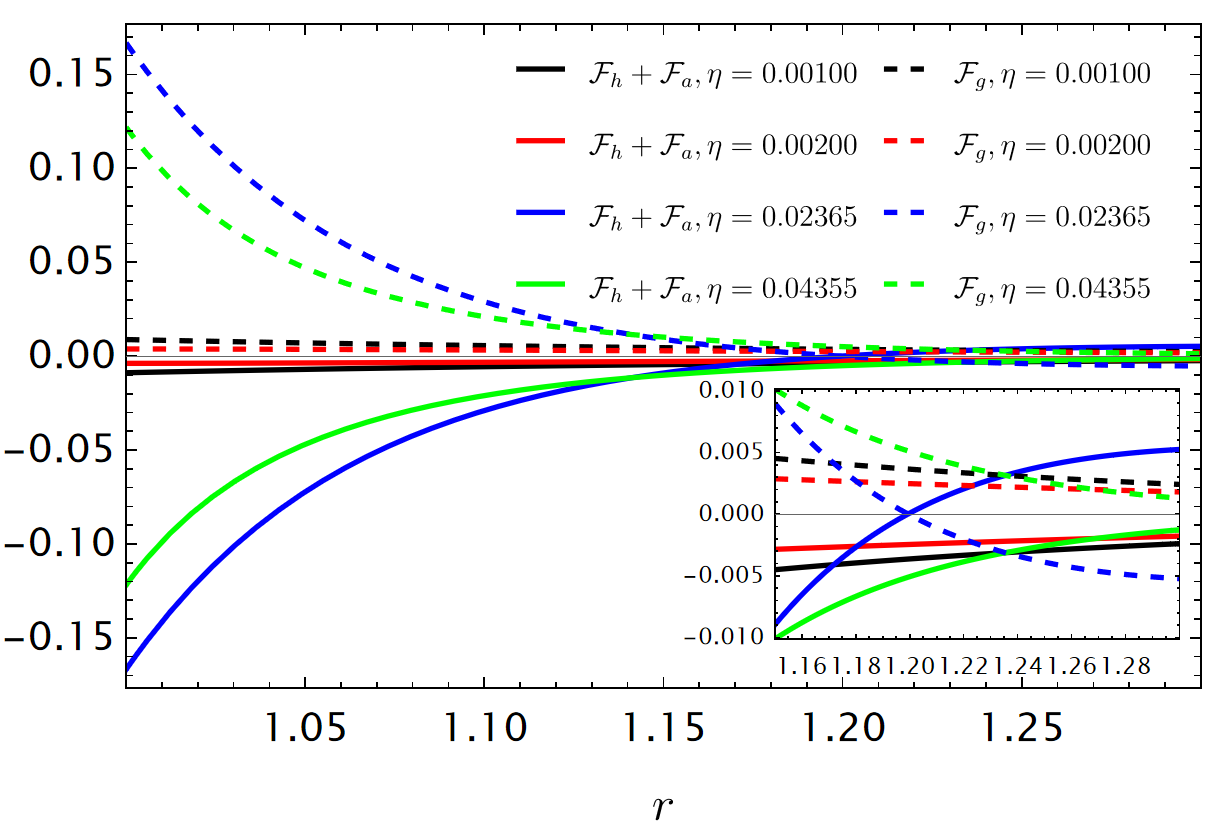}
\includegraphics[width=0.49\linewidth]{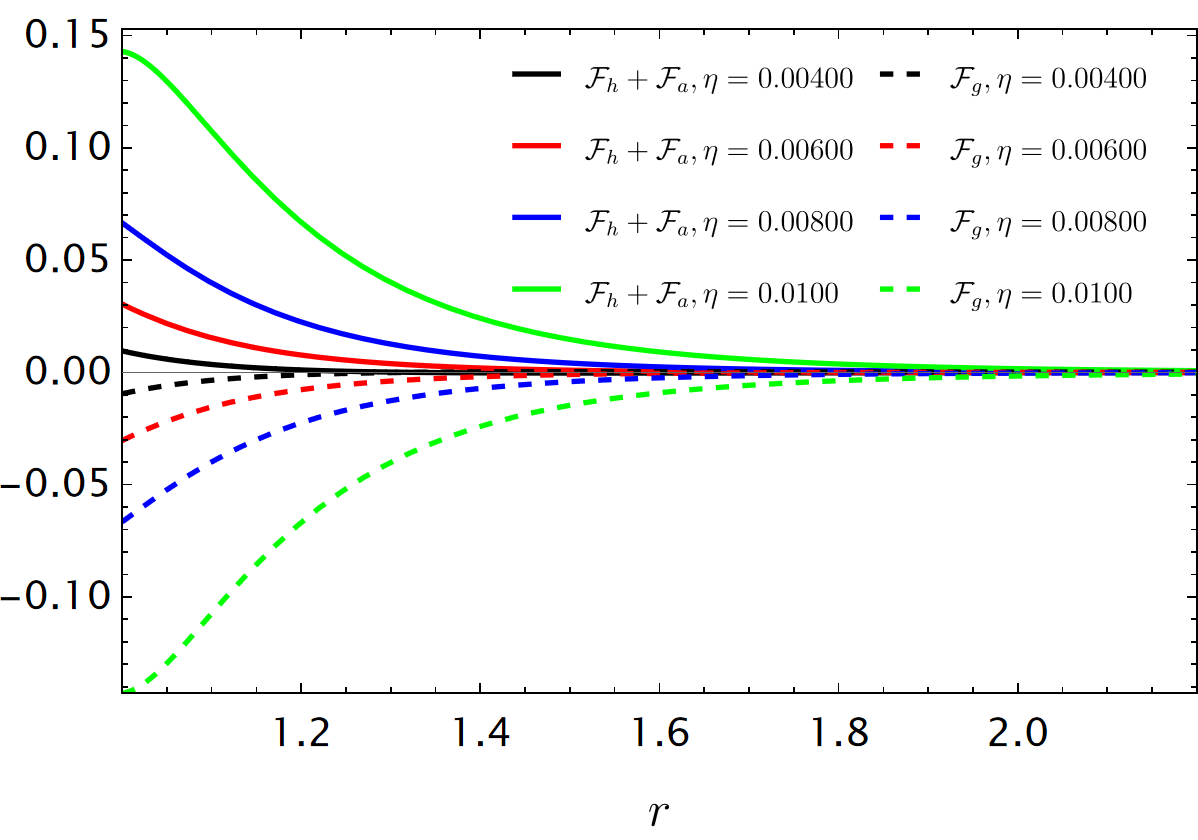}
\caption{TOV force balance for the Casimir-memory wormhole, with $\alpha=\pi^2/720$ and $r_0=1$ fixed. The solid curves represent the combined hydrostatic and anisotropic contribution, $\mathcal{F}_h+\mathcal{F}_a$, while the dashed curves represent the gravitational contribution $\mathcal{F}_g$. In the left panel, we show the regime in which the gravitational contribution is positive near the throat, using $\eta=0.00100$, $0.00200$, $0.02365$, and $0.04355$. The inset magnifies the region where the force contributions approach zero and may change sign away from the throat. In the right panel, we show a complementary regime, with $\eta=0.00400$, $0.00600$, $0.00800$, and $0.01000$, in which the signs of the force contributions are inverted at the throat: the gravitational term becomes negative, while the combined hydrostatic and anisotropic contribution becomes positive. In both panels, the opposite signs of $\mathcal{F}_g$ and $\mathcal{F}_h+\mathcal{F}_a$ reflect the equilibrium condition $\mathcal{F}_g+\mathcal{F}_h+\mathcal{F}_a=0$.} \label{fig:tov_forces}
\end{figure}

Figure~\ref{fig:tov_forces} shows the TOV force balance for the Casimir-memory wormhole. With the definitions adopted here, $\mathcal{F}_g$ denotes the gravitational contribution, while $\mathcal{F}_h+\mathcal{F}_a$ contains the combined effect of the hydrostatic and anisotropic terms. In both panels, these two sectors appear with opposite signs, as required by the equilibrium condition $\mathcal{F}_g+\mathcal{F}_h+\mathcal{F}_a=0$. Therefore, the figure does not represent an unbalanced dynamics of the throat, but rather the internal compensation mechanism that allows the wormhole configuration to remain static. As $r$ increases, all contributions tend to vanish, consistently with the weakening of the effective source and with the asymptotically flat behavior of the geometry.

A relevant feature is that the sign of the gravitational contribution is not fixed throughout the parameter space. In the left panel, the selected values of $\eta$ describe a regime in which $\mathcal{F}_g$ is positive near the throat, while $\mathcal{F}_h+\mathcal{F}_a$ is negative. In this case, the gravitational sector behaves effectively as a repulsive contribution in the throat neighborhood, and the pressure-gradient plus anisotropic sector provides the compensating inward response. In the right panel, however, the signs are inverted at the throat: $\mathcal{F}_g$ becomes negative, while $\mathcal{F}_h+\mathcal{F}_a$ becomes positive. Thus, the gravitational contribution becomes effectively attractive near the throat, and the opening of the wormhole is sustained by the outward action of the hydrostatic and anisotropic terms.

This inversion deserves attention because it shows that the same Casimir-memory source can support the wormhole through different internal mechanisms. A positive $\mathcal{F}_g$ near the throat suggests a regime in which the gravitational contribution, as defined by the TOV decomposition, behaves effectively as a repulsive term. A negative $\mathcal{F}_g$, on the other hand, indicates that the geometry is not being maintained by a locally repulsive gravitational contribution. Instead, the matter anisotropy and the radial pressure gradient become responsible for counterbalancing the attractive tendency and preserving the flare-out geometry. Therefore, the sign of $\mathcal{F}_g$ carries direct information about how the throat is mechanically supported, even though it is not, by itself, a separate condition for traversability.

The radial sign change of the forces is also physically meaningful, but its interpretation depends on the region of parameter space. For values of $\eta$ above the transition scale $\eta_c=\alpha r_0^3$, the near-throat density profile may become locally dominated by the memory-induced contribution before returning to a Casimir-dominated behavior at larger radii, since the memory term decays faster than the ordinary Casimir term. In this case, the sign reversal of the TOV contributions reflects, in part, the radial redistribution of the effective source. For subcritical values of $\eta$, however, the density remains Casimir-dominated throughout the exterior region, and the force inversion is instead associated with the response of the regular redshift sector and of the anisotropic stresses to the memory correction. Thus, the sign of $\mathcal{F}_g$ is controlled not only by the density profile, but by the full combination $-(\rho+p_r)\Phi'$ entering the TOV balance.

Consequently, the change of sign of $\mathcal{F}_g$ should not be interpreted as a pathology of the solution. The geometric viability is still determined by the throat condition, the flare-out condition, the absence of horizons, and asymptotic flatness. Rather, the sign reversal indicates a redistribution of the internal support mechanism. In one regime, the redshift sector contributes as an effectively repulsive gravitational term near the throat, balanced by hydrostatic and anisotropic stresses. In the complementary regime, the gravitational contribution becomes effectively attractive and the opening of the throat is maintained by the outward combined action of pressure gradients and anisotropy. This shows that the memory-corrected Casimir source can reorganize the equilibrium structure of the wormhole, even when the underlying geometry remains traversable.

\section{Phenomenological signatures}
\label{sec:phenomenology}

In this section, we discuss a possible observational diagnostic of the geometry: the shadow associated with unstable photon orbits. Although a realistic comparison with compact-object images requires rotation, accretion physics, and radiative transfer, the static shadow radius provides a first estimate of how the Casimir-memory parameters affect null geodesics in the strong-field region.

For a static and spherically symmetric metric of the form \eqref{eq:mt_metric}, null geodesics in the equatorial plane satisfy an effective radial equation controlled by the redshift function and the shape function. The critical impact parameter is determined by
\begin{equation}
b_{\rm ph}^2=
\frac{r_{\rm ph}^2}{e^{2\Phi(r_{\rm ph})}},
\label{eq:impact_parameter}
\end{equation}
where $r_{\rm ph}$ is the photon-sphere radius. Since $g_{\phi\phi}=r^2$ and $g_{tt}=-e^{2\Phi(r)}$, the photon-sphere condition can be written as
\begin{equation}
\frac{d}{dr}
\left[
r^2 e^{-2\Phi(r)}
\right]_{r=r_{\rm ph}}=0,
\label{eq:photon_sphere_condition}
\end{equation}
or, equivalently,
\begin{equation}
r_{\rm ph}\Phi'(r_{\rm ph})=1.
\label{eq:photon_sphere_simple}
\end{equation}
Thus, the shadow is especially sensitive to the redshift function. In the present construction, $\Phi(r)$ is not arbitrary: it is fixed by the barotropic condition and by regularity at the throat. Therefore, the shadow radius carries information about the same matter-sector balance that determines the wormhole's traversability.

For a distant observer in the asymptotically flat region, the shadow radius is approximately
\begin{equation}
R_{\rm sh}=b_{\rm ph}
=r_{\rm ph}e^{-\Phi(r_{\rm ph})}.
\label{eq:shadow_radius}
\end{equation}
For an observer at a finite radial position $R_o$, one may use
\begin{equation}
R_{\rm sh}\simeq R_o\sin\alpha_{\rm sh},
\qquad
\sin^2\alpha_{\rm sh}
=
\frac{e^{2\Phi(R_o)}}{R_o^2}
r_{\rm ph}^2 e^{-2\Phi(r_{\rm ph})}.
\label{eq:finite_observer_shadow}
\end{equation}

In order to compare the model with EHT bounds, one may express $R_{\rm sh}$ in units of the compact-object mass scale $M$. The parameters $\alpha$, $\eta$, and $r_0$ should then be chosen such that the geometry remains asymptotically flat, satisfies the flare-out condition, and produces a shadow radius inside the observational interval. Since $\eta$ controls the gravitational-memory contribution, its variation changes the throat geometry and the redshift profile, thereby shifting the predicted value of $R_{\rm sh}/M$.

\begin{figure}[!htp]
\centering
\includegraphics[width=0.65\textwidth]{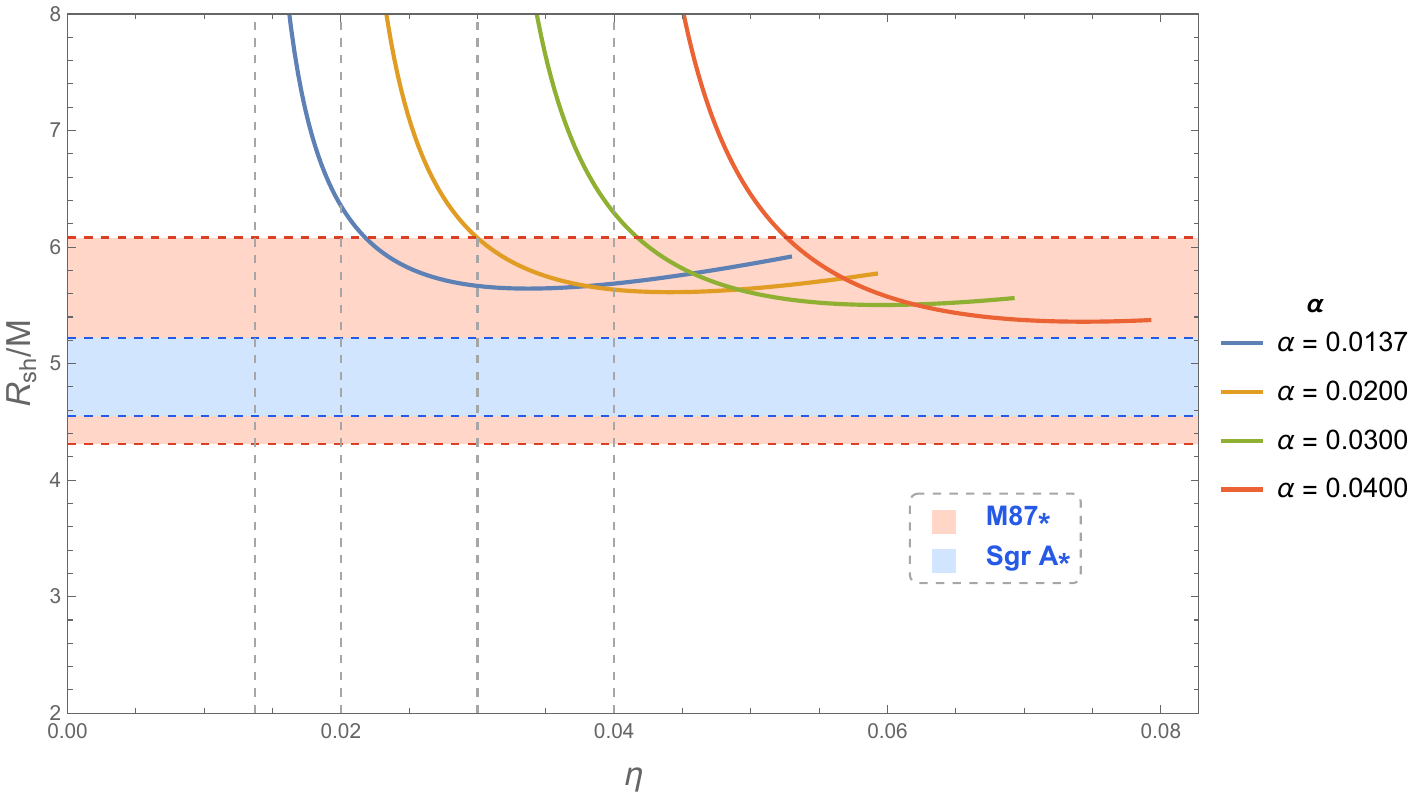}
\caption{Dimensionless shadow radius $R_{\rm sh}/M$ as a function of the memory parameter $\eta$ for different values of the Casimir parameter $\alpha$, with $r_0=1$. The vertical dashed lines indicate the regularity limit $\eta=\alpha$, where the barotropic parameter diverges. The shaded regions correspond to the EHT-inferred ranges for the shadow sizes of M87* and Sgr A*, and are shown as phenomenological benchmarks for comparison with the model predictions.}
\label{fig:shadow_radius}
\end{figure}

As shown in Fig.~\ref{fig:shadow_radius}, the dimensionless shadow radius exhibits a strong dependence on the memory parameter $\eta$ in the vicinity of the regularity limit $\eta=\alpha$, marked by the vertical dashed lines. As $\eta\to\alpha$, the barotropic parameter $\omega=1/[8\pi(\alpha-\eta)]$ diverges, producing a rapid growth of the redshift function and, consequently, of the shadow radius. This behavior explains the steep increase of $R_{\rm sh}/M$ near the asymptotic boundaries of the curves. Physically, this regime corresponds to configurations approaching the limit at which the regularity condition at the throat breaks down.

Moving away from the singular boundary, the shadow radius decreases and enters the observationally favored range inferred by the Event Horizon Telescope. Using the constraints reported in Refs.~\cite{EventHorizonTelescope:2019dse,EventHorizonTelescopeCollaboration_2022}, the corresponding intervals are $4.55\leq R_{\rm sh}/M\leq 5.22$ for Sgr A* and $4.31\leq R_{\rm sh}/M\leq 6.08$ for M87*. For the adopted normalization $r_0=1=2M$, the predicted shadow radii overlap significantly with the M87* interval, whereas they remain systematically above the narrower range inferred for Sgr A*. Interestingly, the EHT-compatible region is found entirely in the sector $\eta>\alpha$, where the effective equation-of-state parameter becomes phantom-like ($\omega<-1$). Nevertheless, the flare-out condition remains satisfied throughout the domain displayed in the figure, ensuring that the corresponding solutions still represent traversable wormholes.

The curves terminate before crossing the entire observational bands because the admissible parameter interval is finite. For each value of $\alpha$, the flare-out condition imposes an upper bound, $\eta < \alpha+\frac{1}{8\pi}$, beyond which the wormhole geometry ceases to exist. Therefore, the endpoints of the curves do not signal any pathology of the shadow itself, but rather the boundary of the geometrically allowed parameter space. Another noteworthy feature is the relatively mild dependence of $R_{\rm sh}/M$ on $\alpha$ once the solutions are sufficiently far from the regularity limit, indicating that the memory parameter $\eta$ plays the dominant role in determining the observable shadow size within this class of Casimir-memory wormholes.

\section{Conclusions}
\label{sec:conclusion}

We have constructed a class of traversable wormhole geometries supported by an effective Casimir source corrected by gravitational memory. In this model, the ordinary negative Casimir contribution is supplemented by a positive memory correction induced by a previous time-dependent gravitational perturbation. The resulting matter profile preserves the exotic character required to sustain a wormhole, but the memory term changes the intensity and the radial distribution of this exoticity.

The corresponding shape function was obtained directly from the Einstein equations and satisfies the throat condition by construction. The flare-out condition restricts the admissible interval of the memory parameter, while the regularity of the redshift function fixes the barotropic parameter once the throat radius and the source parameters are chosen. This leads to two qualitatively distinct regimes. In the Casimir-dominated sector, the matter behaves with a positive barotropic parameter. For $\eta>\alpha r_0^3$, the regular solution enters a phantom-like regime. The transition between these two sectors occurs when the density at the throat changes sign, and this critical case is singular within the constant-barotropic description.

The analysis of the energy conditions confirms that the radial NEC is necessarily violated at the throat, as expected for traversable wormholes. This violation follows directly from the flare-out condition and is independent of the detailed redshift profile. The tangential sector is more sensitive to the redshift gradient, showing that the exotic behavior required by the geometry is anisotropic and mainly concentrated along the radial direction.

We also analyzed curvature properties, embedding diagrams, and the equilibrium of the anisotropic source. The Ricci scalar shows how the memory parameter modifies the local curvature around the throat, while the embedding diagram provides a direct visualization of the spatial opening produced by the shape function. The TOV analysis shows that equilibrium is achieved through the balance among gravitational, hydrostatic, and anisotropic forces, with the memory effect entering indirectly through the density, the shape function, and the pressure sector.

Finally, we investigated the shadow properties of the Casimir-memory wormholes and established a phenomenological connection with EHT observations. Since the photon-sphere condition depends explicitly on the redshift function, the shadow radius probes the same interplay between Casimir energy and gravitational-memory effects that governs the wormhole geometry. Our analysis shows that geometrically admissible wormhole solutions can produce shadow radii compatible with the EHT constraints for M87*, while remaining systematically above the narrower interval inferred for Sgr A* under the normalization $r_0=2M$. We also found that the shadow radius grows sharply as the regularity boundary $\eta=\alpha$ is approached, reflecting the divergence of the barotropic parameter and the associated enhancement of the redshift function. These results suggest that shadow observations may provide a useful observational probe of gravitational-memory effects and their impact on the structure of exotic compact objects, while also highlighting the role of regularity conditions in shaping their observable signatures.

\section*{Acknowledgments}
CRM would like to thank Conselho Nacional de Desenvolvimento Cient\'{i}fico e Tecnol\'ogico (CNPq) for partial financial support, through grant 301122/2025-3. FBL is funded by Funda\c{c}\~ao Cearense de Apoio ao Desenvolvimento Cient\'{i}fico e Tecnol\'ogico (FUNCAP) and by Conselho Nacional de Desenvolvimento Cient\'{i}fico e Tecnol\'ogico (CNPq), grant number 305947/2024-9.

\bibliographystyle{apsrev4-1}
\bibliography{Ref.bib}

@article{EinsteinRosen1935,
  author  = {Einstein, Albert and Rosen, Nathan},
  title   = {The Particle Problem in the General Theory of Relativity},
  journal = {Physical Review},
  volume  = {48},
  pages   = {73--77},
  year    = {1935},
  doi     = {10.1103/PhysRev.48.73}
}

@article{MorrisThorne1988,
  author  = {Morris, Michael S. and Thorne, Kip S.},
  title   = {Wormholes in Spacetime and Their Use for Interstellar Travel: A Tool for Teaching General Relativity},
  journal = {American Journal of Physics},
  volume  = {56},
  pages   = {395--412},
  year    = {1988},
  doi     = {10.1119/1.15620}
}

@article{MorrisThorneYurtsever1988,
  author  = {Morris, Michael S. and Thorne, Kip S. and Yurtsever, Ulvi},
  title   = {Wormholes, Time Machines, and the Weak Energy Condition},
  journal = {Physical Review Letters},
  volume  = {61},
  pages   = {1446--1449},
  year    = {1988},
  doi     = {10.1103/PhysRevLett.61.1446}
}

@article{Visser1989,
  author  = {Visser, Matt},
  title   = {Traversable Wormholes: Some Simple Examples},
  journal = {Physical Review D},
  volume  = {39},
  pages   = {3182--3184},
  year    = {1989},
  doi     = {10.1103/PhysRevD.39.3182}
}

@article{PoissonVisser1995,
  author  = {Poisson, Eric and Visser, Matt},
  title   = {Thin-Shell Wormholes: Linearization Stability},
  journal = {Physical Review D},
  volume  = {52},
  pages   = {7318--7321},
  year    = {1995},
  doi     = {10.1103/PhysRevD.52.7318}
}

@article{HochbergVisser1998,
  author  = {Hochberg, David and Visser, Matt},
  title   = {Dynamic Wormholes, Anti-Trapped Surfaces, and Energy Conditions},
  journal = {Physical Review D},
  volume  = {58},
  pages   = {044021},
  year    = {1998},
  doi     = {10.1103/PhysRevD.58.044021}
}

@article{Teo1998,
  author  = {Teo, Edward},
  title   = {Rotating Traversable Wormholes},
  journal = {Physical Review D},
  volume  = {58},
  pages   = {024014},
  year    = {1998},
  doi     = {10.1103/PhysRevD.58.024014}
}

@article{VisserKarDadhich2003,
  author  = {Visser, Matt and Kar, Sayan and Dadhich, Naresh},
  title   = {Traversable Wormholes with Arbitrarily Small Energy Condition Violations},
  journal = {Physical Review Letters},
  volume  = {90},
  pages   = {201102},
  year    = {2003},
  doi     = {10.1103/PhysRevLett.90.201102}
}

@article{NandiZhangKumar2004,
  author  = {Nandi, Kamal K. and Zhang, Yuan-Zhong and Kumar, K. B. Vijaya},
  title   = {Volume Integral Theorem for Exotic Matter},
  journal = {Physical Review D},
  volume  = {70},
  pages   = {127503},
  year    = {2004},
  doi     = {10.1103/PhysRevD.70.127503}
}

@article{FewsterRoman2005,
  author  = {Fewster, Christopher J. and Roman, Thomas A.},
  title   = {On Wormholes with Arbitrarily Small Quantities of Exotic Matter},
  journal = {Physical Review D},
  volume  = {72},
  pages   = {044023},
  year    = {2005},
  doi     = {10.1103/PhysRevD.72.044023}
}

@article{Casimir1948,
  author  = {Casimir, H. B. G.},
  title   = {On the Attraction Between Two Perfectly Conducting Plates},
  journal = {Proceedings of the Koninklijke Nederlandse Akademie van Wetenschappen},
  volume  = {51},
  pages   = {793--795},
  year    = {1948}
}

@article{CasimirPolder1948,
  author  = {Casimir, H. B. G. and Polder, D.},
  title   = {The Influence of Retardation on the London--van der Waals Forces},
  journal = {Physical Review},
  volume  = {73},
  pages   = {360--372},
  year    = {1948},
  doi     = {10.1103/PhysRev.73.360}
}

@article{BrownMaclay1969,
  author  = {Brown, L. S. and Maclay, G. J.},
  title   = {Vacuum Stress Between Conducting Plates: An Image Solution},
  journal = {Physical Review},
  volume  = {184},
  pages   = {1272--1279},
  year    = {1969},
  doi     = {10.1103/PhysRev.184.1272}
}

@article{Lamoreaux1997,
  author  = {Lamoreaux, S. K.},
  title   = {Demonstration of the Casimir Force in the 0.6 to 6 $\mu$m Range},
  journal = {Physical Review Letters},
  volume  = {78},
  pages   = {5--8},
  year    = {1997},
  doi     = {10.1103/PhysRevLett.78.5}
}

@article{MohideenRoy1998,
  author  = {Mohideen, U. and Roy, Anushree},
  title   = {Precision Measurement of the Casimir Force from 0.1 to 0.9 $\mu$m},
  journal = {Physical Review Letters},
  volume  = {81},
  pages   = {4549--4552},
  year    = {1998},
  doi     = {10.1103/PhysRevLett.81.4549}
}

@article{BordagMohideenMostepanenko2001,
  author  = {Bordag, Michael and Mohideen, Umar and Mostepanenko, V. M.},
  title   = {New Developments in the Casimir Effect},
  journal = {Physics Reports},
  volume  = {353},
  pages   = {1--205},
  year    = {2001},
  doi     = {10.1016/S0370-1573(01)00015-1}
}

@article{Milton2004,
  author  = {Milton, Kimball A.},
  title   = {The Casimir Effect: Recent Controversies and Progress},
  journal = {Journal of Physics A: Mathematical and General},
  volume  = {37},
  pages   = {R209--R277},
  year    = {2004},
  doi     = {10.1088/0305-4470/37/38/R01}
}

@article{Jaffe2005,
  author  = {Jaffe, R. L.},
  title   = {The Casimir Effect and the Quantum Vacuum},
  journal = {Physical Review D},
  volume  = {72},
  pages   = {021301},
  year    = {2005},
  doi     = {10.1103/PhysRevD.72.021301}
}

@article{FullingMiltonParasharRomeoShajeshWagner2007,
  author  = {Fulling, Stephen A. and Milton, Kimball A. and Parashar, Prachi and Romeo, August and Shajesh, K. V. and Wagner, Jef},
  title   = {How Does Casimir Energy Fall?},
  journal = {Physical Review D},
  volume  = {76},
  pages   = {025004},
  year    = {2007},
  doi     = {10.1103/PhysRevD.76.025004}
}

@article{MiltonParasharShajeshWagner2007,
  author  = {Milton, Kimball A. and Parashar, Prachi and Shajesh, K. V. and Wagner, Jef},
  title   = {How Does Casimir Energy Fall? II. Gravitational Acceleration of Quantum Vacuum Energy},
  journal = {Journal of Physics A: Mathematical and Theoretical},
  volume  = {40},
  pages   = {10935--10943},
  year    = {2007},
  doi     = {10.1088/1751-8113/40/35/014}
}

@article{ShajeshMiltonParasharWagner2008,
  author  = {Shajesh, K. V. and Milton, Kimball A. and Parashar, Prachi and Wagner, Jeffrey A.},
  title   = {How Does Casimir Energy Fall? III. Inertial Forces on Vacuum Energy},
  journal = {Journal of Physics A: Mathematical and Theoretical},
  volume  = {41},
  pages   = {164058},
  year    = {2008},
  doi     = {10.1088/1751-8113/41/16/164058}
}

@article{MiltonShajeshFullingParashar2014,
  author  = {Milton, K. A. and Shajesh, K. V. and Fulling, S. A. and Parashar, Prachi},
  title   = {How Does Casimir Energy Fall? IV. Gravitational Interaction of Regularized Quantum Vacuum Energy},
  journal = {Physical Review D},
  volume  = {89},
  pages   = {064027},
  year    = {2014},
  doi     = {10.1103/PhysRevD.89.064027}
}

@article{NazariNouriZonoz2010,
  author  = {Nazari, B. and Nouri-Zonoz, M.},
  title   = {Casimir Effect in a Weak Gravitational Field and the Spacetime Index of Refraction},
  journal = {Physical Review D},
  volume  = {82},
  pages   = {044047},
  year    = {2010},
  doi     = {10.1103/PhysRevD.82.044047}
}

@article{Sorge2019,
  author  = {Sorge, Francesco},
  title   = {Casimir Effect in a Weak Gravitational Field: Schwinger's Approach},
  journal = {Classical and Quantum Gravity},
  volume  = {36},
  pages   = {235006},
  year    = {2019},
  doi     = {10.1088/1361-6382/ab4def}
}

@article{SorgeWilson2019,
  author  = {Sorge, Francesco and Wilson, Justin H.},
  title   = {Casimir Effect in Free-Fall Towards a Schwarzschild Black Hole},
  journal = {Physical Review D},
  volume  = {100},
  pages   = {105007},
  year    = {2019},
  doi     = {10.1103/PhysRevD.100.105007}
}

@article{WilsonSorgeFulling2020,
  author  = {Wilson, Justin H. and Sorge, Francesco and Fulling, Stephen A.},
  title   = {Tidal and Nonequilibrium Casimir Effects in Free Fall},
  journal = {Physical Review D},
  volume  = {101},
  pages   = {065007},
  year    = {2020},
  doi     = {10.1103/PhysRevD.101.065007}
}

@article{KhabibullinKhusnutdinovSushkov2006,
  author  = {Khabibullin, Artem R. and Khusnutdinov, Nail R. and Sushkov, Sergey V.},
  title   = {Casimir Effect in a Wormhole Spacetime},
  journal = {Classical and Quantum Gravity},
  volume  = {23},
  pages   = {627--634},
  year    = {2006},
  doi     = {10.1088/0264-9381/23/3/006}
}

@article{BezerraBezerraMelloKhusnutdinovSushkov2010,
  author  = {Bezerra, V. B. and Bezerra de Mello, E. R. and Khusnutdinov, N. R. and Sushkov, S. V.},
  title   = {Vacuum Polarization of a Massive Scalar Field in a Wormhole Spacetime},
  journal = {Physical Review D},
  volume  = {81},
  pages   = {084034},
  year    = {2010},
  doi     = {10.1103/PhysRevD.81.084034}
}

@article{GarattiniLobo2009,
  author  = {Garattini, Remo and Lobo, Francisco S. N.},
  title   = {Self-Sustained Traversable Wormholes in Noncommutative Geometry},
  journal = {Physics Letters B},
  volume  = {671},
  pages   = {146--152},
  year    = {2009},
  doi     = {10.1016/j.physletb.2008.11.064}
}

@article{Garattini2019,
  author  = {Garattini, Remo},
  title   = {Casimir Wormholes},
  journal = {The European Physical Journal C},
  volume  = {79},
  pages   = {951},
  year    = {2019},
  doi     = {10.1140/epjc/s10052-019-7468-y}
}

@article{SantosMunizMaluf2023,
  author  = {Santos, Alana C. L. and Muniz, C{\'e}lio R. and Maluf, Roberto V.},
  title   = {Yang-Mills Casimir Wormholes in $D=2+1$},
  journal = {Journal of Cosmology and Astroparticle Physics},
  volume  = {2023},
  number  = {09},
  pages   = {022},
  year    = {2023},
  doi     = {10.1088/1475-7516/2023/09/022}
}

@article{GarattiniTzikas2025,
  author  = {Garattini, Remo and Tzikas, Athanasios G.},
  title   = {Rotating Casimir Wormholes},
  journal = {The European Physical Journal C},
  volume  = {85},
  pages   = {336},
  year    = {2025},
  doi     = {10.1140/epjc/s10052-025-14010-6}
}

@article{Sorge:2023sdf,
  author  = {Sorge, Francesco},
  title   = {Gravitational Memory of the Casimir Effect},
  journal = {Physical Review D},
  volume  = {108},
  pages   = {104003},
  year    = {2023},
  doi     = {10.1103/PhysRevD.108.104003}
}

@article{ZeldovichPolnarev1974,
  author  = {Zel'dovich, Ya. B. and Polnarev, A. G.},
  title   = {Radiation of Gravitational Waves by a Cluster of Superdense Stars},
  journal = {Soviet Astronomy},
  volume  = {18},
  pages   = {17--23},
  year    = {1974}
}

@article{BraginskyGrishchuk1985,
  author  = {Braginsky, V. B. and Grishchuk, L. P.},
  title   = {Kinematic Resonance and the Memory Effect in Free-Mass Gravitational Antennas},
  journal = {Soviet Physics JETP},
  volume  = {62},
  pages   = {427--430},
  year    = {1985}
}

@article{Christodoulou1991,
  author  = {Christodoulou, Demetrios},
  title   = {Nonlinear Nature of Gravitation and Gravitational-Wave Experiments},
  journal = {Physical Review Letters},
  volume  = {67},
  pages   = {1486--1489},
  year    = {1991},
  doi     = {10.1103/PhysRevLett.67.1486}
}

@article{Thorne1992,
  author  = {Thorne, Kip S.},
  title   = {Gravitational-Wave Bursts with Memory: The Christodoulou Effect},
  journal = {Physical Review D},
  volume  = {45},
  pages   = {520--524},
  year    = {1992},
  doi     = {10.1103/PhysRevD.45.520}
}

@article{Weinberg1965,
  author  = {Weinberg, Steven},
  title   = {Infrared Photons and Gravitons},
  journal = {Physical Review},
  volume  = {140},
  pages   = {B516--B524},
  year    = {1965},
  doi     = {10.1103/PhysRev.140.B516}
}

@article{HeLysovMitraStrominger2015,
  author  = {He, Temple and Lysov, Vyacheslav and Mitra, Prahar and Strominger, Andrew},
  title   = {BMS Supertranslations and Weinberg's Soft Graviton Theorem},
  journal = {Journal of High Energy Physics},
  volume  = {2015},
  number  = {05},
  pages   = {151},
  year    = {2015},
  doi     = {10.1007/JHEP05(2015)151}
}

@article{Favata2009,
  author  = {Favata, Marc},
  title   = {Nonlinear Gravitational-Wave Memory from Binary Black Hole Mergers},
  journal = {The Astrophysical Journal Letters},
  volume  = {696},
  pages   = {L159--L162},
  year    = {2009},
  doi     = {10.1088/0004-637X/696/2/L159}
}

@article{Favata2010,
  author  = {Favata, Marc},
  title   = {The Gravitational-Wave Memory Effect},
  journal = {Classical and Quantum Gravity},
  volume  = {27},
  pages   = {084036},
  year    = {2010},
  doi     = {10.1088/0264-9381/27/8/084036}
}

@article{BieriGarfinkle2013,
  author  = {Bieri, Lydia and Garfinkle, David},
  title   = {An Electromagnetic Analogue of Gravitational Wave Memory},
  journal = {Classical and Quantum Gravity},
  volume  = {30},
  pages   = {195009},
  year    = {2013},
  doi     = {10.1088/0264-9381/30/19/195009}
}

@article{BieriChenYau2012,
  author  = {Bieri, Lydia and Chen, PoNing and Yau, Shing-Tung},
  title   = {The Electromagnetic Christodoulou Memory Effect and its Application to Neutron Star Binary Mergers},
  journal = {Classical and Quantum Gravity},
  volume  = {29},
  pages   = {215003},
  year    = {2012},
  doi     = {10.1088/0264-9381/29/21/215003}
}

@article{StromingerZhiboedov2016,
  author  = {Strominger, Andrew and Zhiboedov, Alexander},
  title   = {Gravitational Memory, BMS Supertranslations and Soft Theorems},
  journal = {Journal of High Energy Physics},
  volume  = {2016},
  number  = {01},
  pages   = {086},
  year    = {2016},
  doi     = {10.1007/JHEP01(2016)086}
}

@article{PasterskiStromingerZhiboedov2016,
  author  = {Pasterski, Sabrina and Strominger, Andrew and Zhiboedov, Alexander},
  title   = {New Gravitational Memories},
  journal = {Journal of High Energy Physics},
  volume  = {2016},
  number  = {12},
  pages   = {053},
  year    = {2016},
  doi     = {10.1007/JHEP12(2016)053}
}

@article{TolishWald2014,
  author  = {Tolish, Alexander and Wald, Robert M.},
  title   = {Retarded Fields of Null Particles and the Memory Effect},
  journal = {Physical Review D},
  volume  = {89},
  pages   = {064008},
  year    = {2014},
  doi     = {10.1103/PhysRevD.89.064008}
}

@article{LaskyThraneLevinBlackmanChen2016,
  author  = {Lasky, Paul D. and Thrane, Eric and Levin, Yuri and Blackman, Jonathan and Chen, Yanbei},
  title   = {Detecting Gravitational-Wave Memory with LIGO: Implications of GW150914},
  journal = {Physical Review Letters},
  volume  = {117},
  pages   = {061102},
  year    = {2016},
  doi     = {10.1103/PhysRevLett.117.061102}
}

@article{HeisenbergYunesZosso2023,
  author  = {Heisenberg, Lavinia and Yunes, Nicol{\'a}s and Zosso, Jann},
  title   = {Gravitational Wave Memory Beyond General Relativity},
  journal = {Physical Review D},
  volume  = {108},
  pages   = {024010},
  year    = {2023},
  doi     = {10.1103/PhysRevD.108.024010}
}

@article{EventHorizonTelescope:2019dse,
    author = "Akiyama, Kazunori and others",
    collaboration = "Event Horizon Telescope",
    title = "{First M87 Event Horizon Telescope Results. I. The Shadow of the Supermassive Black Hole}",
    eprint = "1906.11238",
    archivePrefix = "arXiv",
    primaryClass = "astro-ph.GA",
    doi = "10.3847/2041-8213/ab0ec7",
    journal = "Astrophys. J. Lett.",
    volume = "875",
    pages = "L1",
    year = "2019"
}

@article{EventHorizonTelescopeCollaboration_2022,
doi = {10.3847/2041-8213/ac6756},
url = {https://dx.doi.org/10.3847/2041-8213/ac6756},
year = {2022},
month = {may},
publisher = {The American Astronomical Society},
volume = {930},
number = {2},
pages = {L17},
author = {Event Horizon Telescope Collaboration and Akiyama, Kazunori and Alberdi, Antxon and Alef, Walter and Algaba, Juan Carlos and Anantua, Richard and Asada, Keiichi and Azulay, Rebecca and Bach, Uwe and Baczko, Anne-Kathrin and Ball, David and Baloković, Mislav and Barrett, John and Bauböck, Michi and Benson, Bradford A. and Bintley, Dan and Blackburn, Lindy and Blundell, Raymond and Bouman, Katherine L. and Bower, Geoffrey C. and Boyce, Hope and Bremer, Michael and Brinkerink, Christiaan D. and Brissenden, Roger and Britzen, Silke and Broderick, Avery E. and Broguiere, Dominique and Bronzwaer, Thomas and Bustamante, Sandra and Byun, Do-Young and Carlstrom, John E. and Ceccobello, Chiara and Chael, Andrew and Chan, Chi-kwan and Chatterjee, Koushik and Chatterjee, Shami and Chen, Ming-Tang and Chen, Yongjun and Cheng, Xiaopeng and Cho, Ilje and Christian, Pierre and Conroy, Nicholas S. and Conway, John E. and Cordes, James M. and Crawford, Thomas M. and Crew, Geoffrey B. and Cruz-Osorio, Alejandro and Cui, Yuzhu and Davelaar, Jordy and De Laurentis, Mariafelicia and Deane, Roger and Dempsey, Jessica and Desvignes, Gregory and Dexter, Jason and Dhruv, Vedant and Doeleman, Sheperd S. and Dougal, Sean and Dzib, Sergio A. and Eatough, Ralph P. and Emami, Razieh and Falcke, Heino and Farah, Joseph and Fish, Vincent L. and Fomalont, Ed and Ford, H. Alyson and Fraga-Encinas, Raquel and Freeman, William T. and Friberg, Per and Fromm, Christian M. and Fuentes, Antonio and Galison, Peter and Gammie, Charles F. and García, Roberto and Gentaz, Olivier and Georgiev, Boris and Goddi, Ciriaco and Gold, Roman and Gómez-Ruiz, Arturo I. and Gómez, José L. and Gu, Minfeng and Gurwell, Mark and Hada, Kazuhiro and Haggard, Daryl and Haworth, Kari and Hecht, Michael H. and Hesper, Ronald and Heumann, Dirk and Ho, Luis C. and Ho, Paul and Honma, Mareki and Huang, Chih-Wei L. and Huang, Lei and Hughes, David H. and Ikeda, Shiro and Impellizzeri, C. M. Violette and Inoue, Makoto and Issaoun, Sara and James, David J. and Jannuzi, Buell T. and Janssen, Michael and Jeter, Britton and Jiang, Wu and Jiménez-Rosales, Alejandra and Johnson, Michael D. and Jorstad, Svetlana and Joshi, Abhishek V. and Jung, Taehyun and Karami, Mansour and Karuppusamy, Ramesh and Kawashima, Tomohisa and Keating, Garrett K. and Kettenis, Mark and Kim, Dong-Jin and Kim, Jae-Young and Kim, Jongsoo and Kim, Junhan and Kino, Motoki and Koay, Jun Yi and Kocherlakota, Prashant and Kofuji, Yutaro and Koch, Patrick M. and Koyama, Shoko and Kramer, Carsten and Kramer, Michael and Krichbaum, Thomas P. and Kuo, Cheng-Yu and Bella, Noemi La and Lauer, Tod R. and Lee, Daeyoung and Lee, Sang-Sung and Leung, Po Kin and Levis, Aviad and Li, Zhiyuan and Lico, Rocco and Lindahl, Greg and Lindqvist, Michael and Lisakov, Mikhail and Liu, Jun and Liu, Kuo and Liuzzo, Elisabetta and Lo, Wen-Ping and Lobanov, Andrei P. and Loinard, Laurent and Lonsdale, Colin J. and Lu, Ru-Sen and Mao, Jirong and Marchili, Nicola and Markoff, Sera and Marrone, Daniel P. and Marscher, Alan P. and Martí-Vidal, Iván and Matsushita, Satoki and Matthews, Lynn D. and Medeiros, Lia and Menten, Karl M. and Michalik, Daniel and Mizuno, Izumi and Mizuno, Yosuke and Moran, James M. and Moriyama, Kotaro and Moscibrodzka, Monika and Müller, Cornelia and Mus, Alejandro and Musoke, Gibwa and Myserlis, Ioannis and Nadolski, Andrew and Nagai, Hiroshi and Nagar, Neil M. and Nakamura, Masanori and Narayan, Ramesh and Narayanan, Gopal and Natarajan, Iniyan and Nathanail, Antonios and Fuentes, Santiago Navarro and Neilsen, Joey and Neri, Roberto and Ni, Chunchong and Noutsos, Aristeidis and Nowak, Michael A. and Oh, Junghwan and Okino, Hiroki and Olivares, Héctor and Ortiz-León, Gisela N. and Oyama, Tomoaki and Özel, Feryal and Palumbo, Daniel C. M. and Paraschos, Georgios Filippos and Park, Jongho and Parsons, Harriet and Patel, Nimesh and Pen, Ue-Li and Pesce, Dominic W. and Piétu, Vincent and Plambeck, Richard and PopStefanija, Aleksandar and Porth, Oliver and Pötzl, Felix M. and Prather, Ben and Preciado-López, Jorge A. and Psaltis, Dimitrios and Pu, Hung-Yi and Ramakrishnan, Venkatessh and Rao, Ramprasad and Rawlings, Mark G. and Raymond, Alexander W. and Rezzolla, Luciano and Ricarte, Angelo and Ripperda, Bart and Roelofs, Freek and Rogers, Alan and Ros, Eduardo and Romero-Cañizales, Cristina and Roshanineshat, Arash and Rottmann, Helge and Roy, Alan L. and Ruiz, Ignacio and Ruszczyk, Chet and Rygl, Kazi L. J. and Sánchez, Salvador and Sánchez-Argüelles, David and Sánchez-Portal, Miguel and Sasada, Mahito and Satapathy, Kaushik and Savolainen, Tuomas and Schloerb, F. Peter and Schonfeld, Jonathan and Schuster, Karl-Friedrich and Shao, Lijing and Shen, Zhiqiang and Small, Des and Sohn, Bong Won and SooHoo, Jason and Souccar, Kamal and Sun, He and Tazaki, Fumie and Tetarenko, Alexandra J. and Tiede, Paul and Tilanus, Remo P. J. and Titus, Michael and Torne, Pablo and Traianou, Efthalia and Trent, Tyler and Trippe, Sascha and Turk, Matthew and van Bemmel, Ilse and van Langevelde, Huib Jan and van Rossum, Daniel R. and Vos, Jesse and Wagner, Jan and Ward-Thompson, Derek and Wardle, John and Weintroub, Jonathan and Wex, Norbert and Wharton, Robert and Wielgus, Maciek and Wiik, Kaj and Witzel, Gunther and Wondrak, Michael F. and Wong, George N. and Wu, Qingwen and Yamaguchi, Paul and Yoon, Doosoo and Young, André and Young, Ken and Younsi, Ziri and Yuan, Feng and Yuan, Ye-Fei and Zensus, J. Anton and Zhang, Shuo and Zhao, Guang-Yao and Zhao, Shan-Shan},
title = {First Sagittarius A* Event Horizon Telescope Results. VI. Testing the Black Hole Metric},
journal = {The Astrophysical Journal Letters}
}

@article{Hassan2023,
   author = {Zinnat Hassan and Sayantan Ghosh and P. K. Sahoo and V. Sree Hari Rao},
   doi = {10.1007/s10714-023-03139-y},
   issn = {0001-7701},
   issue = {8},
   journal = {General Relativity and Gravitation},
   month = {8},
   pages = {90},
   publisher = {Springer},
   title = {GUP corrected Casimir wormholes in f(Q) gravity},
   volume = {55},
   url = {https://link.springer.com/10.1007/s10714-023-03139-y},
   year = {2023}
}

@article{Hassan2022,
   author = {Zinnat Hassan and Sayantan Ghosh and P. K. Sahoo and Kazuharu Bamba},
   doi = {10.1140/epjc/s10052-022-11107-0},
   issn = {1434-6052},
   issue = {12},
   journal = {The European Physical Journal C},
   month = {12},
   pages = {1116},
   publisher = {Institute for Ionics},
   title = {Casimir wormholes in modified symmetric teleparallel gravity},
   volume = {82},
   url = {https://link.springer.com/10.1140/epjc/s10052-022-11107-0},
   year = {2022}
}

@article{Santos2024,
   author = {A. C. L. Santos and R. V. Maluf and C. R. Muniz},
   doi = {10.1016/j.aop.2024.169775},
   issn = {00034916},
   journal = {Annals of Physics},
   month = {8},
   pages = {169775},
   title = {Generating 4-dimensional Wormholes with Yang-Mills Casimir Sources},
   volume = {469},
   url = {http://arxiv.org/abs/2405.08774 http://dx.doi.org/10.1016/j.aop.2024.169775},
   year = {2024}
}

\end{document}